\documentclass[aps,prd,nofootinbib,twocolumn,superscriptaddress,preprintnumbers,letterpaper]{revtex4}

\usepackage{amssymb}
\usepackage{amsmath}
\usepackage{epsfig}
\usepackage[colorlinks=true
,urlcolor=blue
,anchorcolor=blue
,citecolor=blue
,filecolor=blue
,linkcolor=red
,menucolor=blue
,linktocpage=true
,pdfproducer=medialab
,pdfa=true
]{hyperref}
\usepackage{comment}
\usepackage{color}
\usepackage{cleveref}
\usepackage{multirow}
\usepackage{capt-of}
\usepackage{siunitx}
\usepackage{graphicx}
\usepackage{gensymb}
\usepackage[caption=false]{subfig}
\usepackage{slashed}
\usepackage{tikz-feynman}
\usepackage{tabularx}
\usepackage{braket}
\usepackage{ulem}

\makeatletter
\def\simgt{\mathrel{\lower2.5pt\vbox{\lineskip=0pt\baselineskip=0pt
           \hbox{$>$}\hbox{$\sim$}}}}
\def\simlt{\mathrel{\lower2.5pt\vbox{\lineskip=0pt\baselineskip=0pt
           \hbox{$<$}\hbox{$\sim$}}}}
\makeatother

\newcommand\cB{\mathcal{B}}
\newcommand\cL{\mathcal{L}}

\preprint{MIT-CTP/4970}

\begin{document}

\title{
Complementarity for Dark Sector Bound States
}

\author{Gilly Elor}
\email{gelor@uw.edu}
\affiliation{Institute of Theoretical Science, University of Oregon, Eugene, OR 97403, U.S.A.}
\affiliation{Department of Physics, Box 1560, University of Washington, Seattle, WA 98195, U.S.A.}

\author{Hongwan Liu}
\email{hongwan@mit.edu}
\affiliation{Center for Theoretical Physics, Massachusetts Institute of Technology, Cambridge, MA 02139, U.S.A.}

\author{Tracy R. Slatyer}
\email{tslatyer@mit.edu}
\affiliation{Center for Theoretical Physics, Massachusetts Institute of Technology, Cambridge, MA 02139, U.S.A.}

\author{Yotam Soreq}
\email{soreqy@mit.edu}
\affiliation{Center for Theoretical Physics, Massachusetts Institute of Technology, Cambridge, MA 02139, U.S.A.}

\begin{abstract} 
We explore the possibility that bound states involving dark matter particles could be detected by resonance searches at the LHC, and the generic implications of such scenarios for indirect and direct detection. We demonstrate that resonance searches are complementary to mono-jet searches and can probe dark matter masses above 1\,TeV with current LHC data. We argue that this parameter regime, where the bound-state resonance channel is the most sensitive probe of the dark sector, arises most naturally in the context of non-trivial dark sectors with large couplings, nearly-degenerate dark-matter-like states, and multiple force carriers. The presence of bound states detectable by the LHC implies a minimal Sommerfeld enhancement that is appreciable, and potentially also radiative bound state formation in the Galactic halo, leading to large signals in indirect searches. We calculate these complementary constraints, which favor either models where the bound-state-forming dark matter constitutes a small fraction of the total density, or models where the late-time annihilation is suppressed at low velocities or late times. We present concrete examples of models that satisfy all these constraints and where the LHC resonance search is the most sensitive probe of the dark sector.

\end{abstract}

\maketitle

\section{Introduction}
\label{sec:AltIntroduction}

The existence of dark matter~(DM) is well-established by observations of its gravitational effects. However, the particle nature of DM is still very much a mystery, despite the ongoing efforts of many complementary experimental searches. Constraints set by XENON~\cite{Aprile:2017iyp}, LUX~\cite{Akerib:2016vxi} and PandaX~\cite{Cui:2017nnn} have strongly ruled out generic DM candidates that interact in a spin-independent manner through a $Z$-exchange, and are now starting to probe Higgs-mediated interactions ({\it e.g.}~\cite{Escudero:2016gzx}).
These direct detection experiments are complemented by dark sector searches at colliders. 
The main DM search strategy at the Large Hadron Collider~(LHC) is based on missing transverse momentum (MET) balanced by a jet, electroweak~(EW) gauge boson or Higgs, known generically as mono-$X$ searches. 
Searches for dijet or dilepton resonances, while not directly probing the existence of DM, can also effectively constrain models where a mediator particle is responsible for interactions between the Standard Model~(SM) and a ``dark sector'' containing the DM, limiting the parameter space for the mediator. 
Finally, indirect searches for DM annihilation or decay to SM particles, as well as the well-measured relic abundance of the DM, set powerful limits on the strength and nature of the interaction of DM with the SM. Any model of DM must successfully contend with all of these constraints.

With no hint yet of what the dark sector may look like, we might look to the SM for clues as to its possible composition and structure. 
In this light, we should not be surprised to find bound states in the dark sector; after all, bound states are ubiquitous in the SM, and even the simplest dark sector models with a DM candidate and a force carrier can potentially support the existence of bound states. 
Dark sector bound states, much like QCD bound states, may be produced when a pair of heavy dark sector particles are produced close to their kinematic threshold and have a sufficiently strong attractive interaction between them. The subsequent decay of these bound states into lighter SM particles can lead to distinctive signatures at the LHC. This strategy has been studied in the context of bound states formed by supersymmetric~(SUSY) particles, and has been shown to be a potential search channel at the LHC~\cite{Drees:1993uw,Martin:2008sv,Kats:2009bv,Kats:2012ym}, capable of probing regions of parameter space where traditional searches are challenging.

Dark sector bound states and their potential collider signatures have been studied extensively in the literature. 
Bound states formed from weakly-interacting massive particles~(WIMPs) that are charged under the SM SU(2)$_L$ $\times$ U(1)$_Y$ gauge group or non-SM forces, known as WIMPonium~\cite{Shepherd:2009sa}, can be detected at the LHC through resonant decays into a pair of leptons, provided the coupling to the mediator which supports the bound state is large enough. 
Other model-specific dark sector bound state collider searches that have been proposed include searches for higgsino bound states in $\lambda$-SUSY and bound states within the self-interacting DM framework~\cite{Tsai:2015ugz}; DM bound states in a U(1) vector portal model decaying into multilepton final states, which can be searched for at $B$-factories~\cite{An:2015pva}; and a Higgs portal model with decays to electrons which can be searched for at the LHC \cite{Bi:2016gca}. Mono-photon searches at lepton colliders can also potentially be used to probe the full resonance structure of the dark sector~\cite{Hochberg:2017khi}. 
However, the large couplings typically required for detectable bound states often predict large signals in direct detection experiments, especially if the light force carrier responsible for the bound state formation also couples to the SM; likewise, in this light-mediator regime, searches for the mediator are often a more promising dark-sector discovery channel than searches for the bound states~\cite{Han:2007ae}.

In this paper, we broadly explore the challenges of building a dark sector model which can be discovered through the production of a bound state at the LHC, in light of the current stringent and complementary experimental constraints. 
Direct detection limits can be evaded in models with TeV-mass DM if the DM candidate only has an off-diagonal coupling to the SM that couples the DM, the mediator and a heavier dark sector state, so that at tree-level, the DM only scatters into this heavier state when interacting with the SM~\cite{Smith:2001hy, TuckerSmith:2004jv}. 
At the LHC, dark sector particles can be produced on their kinematic threshold and form a bound state $\mathcal{B}$, which can subsequently undergo annihilation decay into a pair of SM leptons, showing up as a dilepton resonance at the LHC.\footnote{Di-jets are also a plausible search strategy, but the backgrounds and triggers make this much more challenging to explore.}

We will show that in models where the mediator between the SM and the dark sector couples to two different states in the dark sector, it is possible to arrange for such a resonance to occur and have a substantial branching ratio into SM leptons. 
In these scenarios, searches for a dilepton resonance from $\mathcal{B}$ are complementary to the existing mono-$X$ and vector resonance searches that are already deployed for dark sector searches at the LHC, with the ability to probe higher mass scales for the mediator and DM. Since $\mathcal{B}$ can have the same quantum numbers as the SM mediator, we explore the importance of mixing between bound states and mediator particles with equal quantum numbers and similar masses.

Models with bound states that are detectable at the LHC can also possess large indirect signals, as the long-range potential implied by the existence of bound states generically enhances the annihilation cross section for slow-moving DM particles, and the bound state formation and decay can also serve as an annihilation  channel. We will study the constraints from indirect detection and cosmology that result from considering these effects.

The rest of the paper is structured as follows. 
In Sec.~\ref{sec:PhenoBound}, we will make some remarks on the general features of dark sector models where the bound-state resonance search is the most sensitive channel.
We will discuss why the bound-state resonance search is complementary to the current dark sector search strategies used by the LHC experiments for such models, and discuss their general phenomenology in direct, indirect and collider DM searches. 
In Sec.~\ref{sec:darksector} we will lay out some specific models containing bound states in the dark sector and study their phenomenology. We will first discuss the MSSM in the pure wino/higgsino limit, which already meets some of the criteria needed for a successful model with bound states, although the production rate at the 13\,TeV LHC is too small for detection. We will then discuss two vector portal models which realize the requirements needed for a viable dark sector with bound states to be probed by the LHC. In Sec.~\ref{sec:experimentalConstraints} we compute and discuss the potential experimental signatures of these models.
Our conclusion will then follow in Sec.~\ref{sec:dis}. 

\section{Phenomenology of Bound States}
\label{sec:PhenoBound}

The existence of DM bound states has implications for the phenomenology of the dark sector, and for its signatures in direct, indirect and collider searches. In this section, we consider the circumstances under which collider searches for bound states can probe otherwise unexplored regions of DM parameter space. Aside from these searches, DM bound states with long lifetimes have also recently been shown to have potentially interesting implications for neutrino experiments~\cite{Grossman:2017qzw}.

As we will show, models where bound-state resonance searches at the LHC probe new regions of parameter space are most easily realized in the presence of several common features:
\begin{enumerate}
\item DM couples to at least two distinct force carriers; one of these, $Y$, is light and mediates the bound state formation, while the other, $V$, is heavier and couples appreciably to the SM. The constraints from LHC resonance searches of the bound state are most competitive when the SM mediator $V$ is heavier than twice the DM mass;
\item the coupling of the DM to the light mediator, which we denote $\alpha_\cB \equiv g_\cB^2/4\pi$, should be fairly large, as the bound state production rate is proportional to the third power of this parameter;
\item decay of $s$-wave bound states with the same spin as the heavy mediator into a pair of light mediators is suppressed, so that decays through the heavier mediator into two SM fermions dominate; and
\item the relevant spin-independent direct detection cross section is suppressed, \textit{e.g.} by loops, by momentum-dependent factors, or by small couplings. This is particularly easy to achieve in models where the DM is part of a multiplet with small mass splittings, and the heavy mediator has an off-diagonal coupling to the mass eigenstates, so that elastic scattering off nuclei occurs only at one-loop level. 
An alternate approach to this criterion would be to consider flavor-dependent couplings between $V$ and the quarks. 
\end{enumerate}
The mono-$X$ process, resonant production of the mediator $V$ and the resonant production of the bound state $\mathcal{B}$ are the main collider signatures of this general setup, and are depicted in Fig.~\ref{fig:feynmanMonoXAndResonance}. When discussing generic models, we will denote the heavy mediator as $V$ and its mass by $m_V$, and the light mediator by $Y$ and its mass as $m_Y$ (for ``Yukawa''). 
In the example models we present, $V$ will be a vector in all cases, but $Y$ can be either a scalar or vector. 
In principle, $V$ could also be a scalar (or a scalar bound state can mix directly with the Higgs sector \cite{Bi:2016gca}), but we will leave the analysis of such scenarios to future work; as we will see, a vector mediator facilitates a sizable production cross section and a large branching ratio to leptons, while evading direct detection bounds.

\begin{table}
    \begin{center}
    \begin{tabular}{c}
    	\begin{tikzpicture}
    		\begin{feynman}
    			\vertex (a1) {\(q\)};
    			\vertex[below right=0.75cm and 0.4375cm of a1] (jet);
    			\vertex[above right=0.75cm and 0.75cm of jet] (jetEnd) {\(X\)};
    			\vertex[below right=1.2cm and 0.7cm of a1] (c1);
                \vertex[left=0.2cm of c1] (gqlabel) {\(g_q\)};
    			\vertex[below left =1cm and 0.5cm of c1] (a2) {\(\overline{q}\)};
    			\vertex[right=1.5cm of c1] (c2);
                \vertex[right=0.2cm of c2] (gchilabel) {\(g_\chi\)};
    			\vertex[above right=1cm and 0.5cm of c2] (b1) {\(\chi\)};
    			\vertex[below right=1cm and 0.5cm of c2] (b2) {\(\overline{\chi}\)};
    			\diagram*{
    				(a1) -- [fermion] (jet);
    				(jet) -- [fermion] (c1);
    				(jetEnd) -- [scalar] (jet);
    				(c1) -- [fermion] (a2); 
    				(c1) -- [boson, edge label'=\(V^{(*)}\)] (c2);
    				(c2) -- [fermion] (b1);
    				(b2) -- [fermion] (c2);
    			};
    		\end{feynman}
    	\end{tikzpicture} \\
    	\begin{tikzpicture}
    		\begin{feynman}
    			\vertex (a1) {\(q\)};
    			\vertex[below right=1.2cm and 0.7cm of a1] (c1);
    			\vertex[below left =1cm and 0.5cm of c1] (a2) {\(\overline{q}\)};
    			\vertex[right=1.5cm of c1] (c2);
    			\vertex[above right=1cm and 0.5cm of c2] (b1) {\(q, \ell^-\)};
    			\vertex[below right=1cm and 0.5cm of c2] (b2) {\(\overline{q}, \ell^+\)};
    			\diagram*{
    				(a1) -- [fermion] (c1);
    				(c1) -- [fermion] (a2); 
    				(c1) -- [boson, edge label'=\(V\)] (c2);
    				(c2) -- [fermion] (b1);
    				(b2) -- [fermion] (c2);
    			};
    		\end{feynman}
    	\end{tikzpicture} \\
        \begin{tikzpicture}
          \begin{feynman}
            \vertex (a1) {\(q\)};
            \vertex[below right=1.2cm and 0.7cm of a1] (c1);
            \vertex[below left =1cm and 0.5cm of c1] (b1) {\(\overline{q}\)};
            \vertex[right=0.9cm of c1] (c3);
            \vertex[above right=0.7cm and 0.5cm of c3] (a2);
            \vertex[above =0.1cm of a2] (alphaBlabel) {\(\alpha_{\mathcal{B}}\)};
            \vertex[right=0.7cm of a2] (a3);
            \vertex[right=0.3cm of a3] (a4) {\( \)};
            \vertex[below right=0.7cm and 0.5cm of c3] (b2);
            \vertex[right=0.7cm of b2] (b3);
            \vertex[right=0.3cm of b3] (b4) {\( \)};
            \vertex[below right=0.7cm and 0.8cm of a4] (c4) {\(\cdots\)};
            \vertex[right=0.5cm of a4] (d1);
            \vertex[right=0.8cm of d1] (d2);
            \vertex[right=0.5cm of b4] (e1);
            \vertex[right=0.8cm of e1] (e2);
            \vertex[below right=0.7cm and 0.5cm of d2] (f1);
            \vertex[right=0.9cm of f1] (g1);
            \vertex[above right=1cm and 0.5cm of g1] (h1) {\(q, \ell^-\)};
    		\vertex[below right=1cm and 0.5cm of g1] (h2) {\(\overline{q}, \ell^+\)};

            \diagram*{
                (a1) -- [fermion] (c1);
                (c1) -- [fermion] (b1);
                (c1) -- [boson, edge label'=\(V^{(*)}\)] (c3);
                (c3) -- [fermion] (a2);
                (b2) -- [fermion] (c3);
                (a2) -- [fermion] (a3);
                (a3) -- (a4);
                (b3) -- [fermion] (b2);
                (b4) -- (b3);
                (a2) -- [scalar, edge label=\(Y\)] (b2);
                (a3) -- [scalar] (b3);
                (d2) -- [scalar] (e2);
                (d1) -- [fermion, edge label=\(\chi\)] (d2);
                (e2) -- [fermion, edge label=\(\overline{\chi}\)] (e1);
                (d2) -- [fermion] (f1);
                (f1) -- [fermion] (e2);
                (f1) -- [boson, edge label'=\(V^{(*)}\)] (g1);
                (g1) -- [fermion] (h1);
                (h2) -- [fermion] (g1);
            };
            
            \draw[decoration={brace}, decorate] (a4.north east) -- (b4.south east) node[pos=0.5, right] {\(\cB\)};
          \end{feynman}
        \end{tikzpicture}
    \end{tabular}
    \end{center}
    
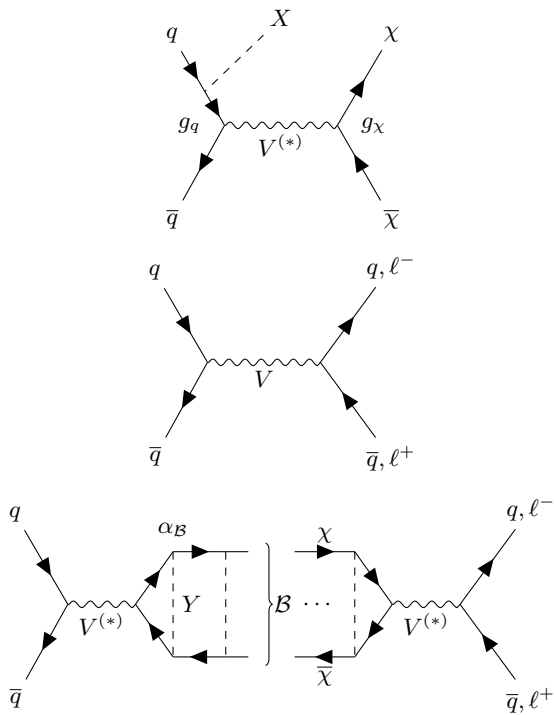
\captionof{figure}{Feynman diagrams for relevant dark sector processes at colliders. These processes are (top) the mono-$X$ process, (middle) the resonant production of $V$ decaying into a pair of jets or leptons, and (bottom) the resonant production of $\cB$, subsequently undergoing a similar decay. The coupling of the mediator between the dark sector and the SM to quarks ($g_q$) and to the DM ($g_\chi$), as well as the coupling responsible for the Yukawa potential that forms the bound state $\mathcal{B}$ ($\alpha_{\mathcal{B}}\equiv g_\cB^2/4\pi$) are shown. In our models, $V$ is always a vector, while $Y$ can be either a scalar or a vector.}
    \label{fig:feynmanMonoXAndResonance}
\end{table}

Many of the earlier works in the literature on bound states exhibit some of these features. 
Both~\cite{Shepherd:2009sa} and~\cite{Tsai:2015ugz} introduce an additional mediator to support the bound state formed from DM charged under the EW gauge group, so that the couplings between the DM to the light mediator can be made large. An additional mediator was also introduced in~\cite{Grossman:2017qzw} to alleviate the tension between a suitably light mediator that can support a bound state and the need for a massive enough SM mediator that can decay into electron pairs. In~\cite{An:2015pva}, direct detection limits are avoided by having sub-GeV DM. 
Furthermore, there is only one vector boson to mediate both the bound state formation and the interaction with the SM, at the cost of allowing the bound state to decay into 4- or 6-lepton final states. This is an important signature in $B$-factories for DM with a mass on the order of a GeV~\cite{An:2015pva}. In principle, this scenario can be probed at the LHC by multi-lepton searches, or by di-photon searches where two $e^+e^-$ pairs are detected as fake photons~\cite{Bi:2016gca}. However, multi-lepton signatures turn out to be relatively unimportant for the kinetic mixing models that we will study later.

We will demonstrate that the characteristics listed above can be achieved in Higgsed dark-sector models, with a vector portal between the dark sector and the SM. But before we give examples of such models, we will first discuss each of these criteria in more detail.

\subsection{General Model Building Considerations}
\label{sec:modelBuilding}

The existence of DM bound states in a Yukawa potential with range $1/m_Y$ is only possible if~\cite{PhysRevA.1.1577, An:2015pva}:
\begin{align}
	\label{eqn:boundStateRequirement}
	\frac{\alpha_\cB \, m_\chi}{m_Y} > 1.68,
\end{align}
where $m_\chi$ is the DM mass. Thus, the presence of a bound state supported by scalar or vector exchange requires a relatively light force carrier -- certainly lighter than the dark matter itself, for weak couplings. For more complex dark sectors with potentials that couple multiple two-particle states (\textit{e.g.} the neutralino sector of supersymmetric models), the details of this criterion may be modified, but it is still generically true that there must be a force with range longer than the Bohr radius of the bound state, i.e. there should be at least one mediator with $m_Y \lesssim \alpha_\mathcal{B} m_\chi$.

If this force carrier is also the mediator between the DM and the SM, then searches for the force carrier will generally offer a more accessible probe of dark-sector physics than searches for the heavier DM, both because the force carrier is lighter and because it couples directly to SM particles (see \textit{e.g.}~\cite{Han:2007ae}). 
This leads us to consider models where there are at least two distinct particles that couple to the DM, one which has appreciable interactions with the SM (and can be heavier than the DM itself), and the other of which mediates the bound state formation and so must be light.

One alternative to this structure is the case where the DM is charged under the SM SU(2)$_L$ EW gauge group, and the photon, $W$ and/or $Z$ support the bound state; this is possible, for example, for bound states consisting of neutralinos and/or charginos~\cite{Asadi:2016ybp}. 
However, as we will show later in this work, at present-day colliders the production rate for such EW bound states is undetectably low. 

Returning to dark-sector models with at least two mediators, the presence of the light mediator has some immediate implications. 
First, the DM will generically annihilate into the light mediators. If these mediators are absolutely stable, they will constitute some fraction of the DM relic density, which must be sufficiently small; if they are below the $\sim$ MeV scale in mass, they may be constrained by limits on the number of effective relativistic degrees of freedom in the early universe (\textit{e.g.}~\cite{Nollett:2013pwa,Nollett:2014lwa}). 
We will generally assume that these mediators decay through some small mixing with the SM, on timescales less than one second, so that they do not affect Big Bang nucleosynthesis; in this case, while the coupling can be made small enough that these mediators do not contribute substantially to collider signals and direct detection, indirect detection constraints from this annihilation channel must be considered.

Resonance searches for bound states will typically become difficult when there is a significant branching ratio of the bound state into dark sector states (while the dark sector states could decay promptly to SM particles as in Ref.~\cite{An:2015pva}, for the heavy DM models considered in this paper, which are most relevant for LHC searches, we expect this signature to be relatively unimportant, as we will explain later). Thus, a model where collider searches for bound state resonances are effective must ensure that the branching ratio of the bound state (formed at a collider) into two light mediators is small relative to the decay of the bound state via the off-shell heavy mediator into a pair of SM particles. 
One way that this can be achieved is if both the heavy and light mediators are vectors, then suppression of the spin-1 bound state decay to two light mediators is automatic: charge parity symmetry forbids the decay of a spin-1 $s$-wave bound state into two vectors, so any decays into dark sector vectors must be a 3-body process.
In fact, decays into any number of the light mediators can be completely forbidden if the bound state is formed not from a particle-antiparticle pair, but from two different fermions in the bound state with nontrivial quantum numbers, which cannot be conserved if the bound state could decay into states containing only light mediators. This behavior is natural in cases where the mediator $V$ couples off-diagonally to the multiplet containing the DM. Models of this type have additional advantages in evading constraints from direct detection. 

Note that if the mediator to the SM is a vector, then the bound states formed at the LHC by resonant production will dominantly be spin-1 $s$-wave states; if the mediator is a scalar, they will instead dominantly be spin-0 $s$-wave states. The spin and angular momentum of the bound states determine their possible decays.

\subsection{Vector-Bound State Mixing}
\label{sec:VBMixing}

When $V$ and $\mathcal{B}$ have similar masses or the coupling between $V$ and the constituents of $\mathcal{B}$ is large, significant mixing can occur between the two states if they have the same quantum numbers. Both the $V$-resonance and $\mathcal{B}$-resonance diagrams in Fig.~\ref{fig:feynmanMonoXAndResonance}, together with higher order diagrams with more inter-conversions between $V$ and $\mathcal{B}$, need to be re-summed. The new mass eigenstates that result from the mixing have masses and widths that are shifted with respect to their unmixed values by an amount determined by the strength of the mixing.

The formalism that accounts for the mixing was used to study $Z$-toponium mixing \cite{Kuhn:1985eu,Kuhn:1987ty,Gusken:1985tf,Hall:1985jf,Franzini:1987jw}, and more recently to study Higgs-stoponium mixing \cite{Bodwin:2016whr}. The mixing shifts the masses and widths of the unmixed states, denoted $V_0$ and $\mathcal{B}_0$, to new values given by the eigenvalues of the following mass matrix:
\begin{alignat}{1}
    \mathcal{M} = \begin{pmatrix}
        m_{V,0}^2 - i m_{V,0} \Gamma_{V,0}(s) & -f \\
        -f & m_{\mathcal{B},0}^2 - i m_{\mathcal{B},0} \Gamma_{\mathcal{B},0}(s)
    \end{pmatrix},
\end{alignat}
where all masses and widths are for the unmixed states, and $f$ is a model-dependent parameter determined by the coupling between $V_0$ and $\mathcal{B}_0$.

If $f$ is small compared to the difference in the diagonal entries (see Eq.~(\ref{eqn:tan2theta}) below), the final mixed states $V$ and $\mathcal{B}$ are approximately their respective initial unmixed states, up to higher order corrections. The width of $V_0$ should be evaluated at the appropriate energy scale $\sqrt{s}$ at which the final mixed resonances $V$ or $\mathcal{B}$ are produced; this scale dependence is important especially when $m_\mathcal{B}$ lies below the $\chi \overline{\chi}$ open production threshold while $m_{V,0}$ lies above it. 
The width of $\mathcal{B}_0$ should not include decays through mixing with the $V$: such effects are exactly what the mixing accounts for. For the kinetic mixing models that we will consider later, we take $\Gamma_{\mathcal{B},0} = 0$, since the dark sector particles do not have any tree-level coupling to the SM, and the unmixed width of the bound state excluding mixing into the SM is always much smaller than $\Gamma_{V,0}$.

After mixing, the mixed mass eigenstates are rotated by a complex mixing angle $\theta$ with respect to the unmixed states, and the masses and widths are shifted by \cite{Hall:1985jf,Kuhn:1987ty}
\begin{alignat}{1}
    Q_V &= Q_{V,0} \cos^2 \theta + Q_{\mathcal{B},0} \sin^2 \theta + f \sin 2 \theta, \nonumber \\
    Q_\mathcal{B} &= Q_{V,0} \sin^2 \theta + Q_{\mathcal{B},0} \cos^2 \theta - f \sin 2 \theta,
    \label{eqn:massMixing}
\end{alignat}
where $Q_j \equiv m_j^2 - i m_j \Gamma_j$, with
\begin{alignat}{1}
    \tan 2 \theta = \frac{2f}{Q_{V,0} - Q_{\mathcal{B},0}}.
    \label{eqn:tan2theta}
\end{alignat}
The rotated mass eigenstate $\mathcal{B}$ therefore develops a coupling to the SM through its $V_0$ component.

When the mixed masses $m_V$ and $m_\mathcal{B}$ are nearly equal, a resonance search for each individual mass eigenstate becomes impossible, since the $s$-channel diagrams with intermediate $V$- and $\mathcal{B}$-states interfere with each other, and the end result is a cross section that may not have a Breit-Wigner form. However, if $\theta$ is small, Eq.~(\ref{eqn:massMixing}) shows that the mixed mass eigenstates are separated by $\Delta m^2 \sim 4f \text{Re}(\theta)$, where $\text{Re}(\theta)$ is the real part of $\theta$. Furthermore, the shift in the masses defined by Eq.~(\ref{eqn:massMixing}) and~(\ref{eqn:tan2theta}) always results in a mass eigenstate that is lighter than both $m_{V,0}$ and $m_{\mathcal{B},0}$, and is therefore always strictly below the threshold for open production of $\chi \overline{\chi}$. These two facts can ensure that the lighter resonance is always narrow, as it cannot decay into $\chi \overline{\chi}$, and is always well-separated from the heavier resonance. We have checked that this is always the case for the models that we consider later.

Finally, in the limit of small $\theta$, this mixing procedure gives a final decay width $\Gamma_\mathcal{B}$ that agrees with the perturbative calculation to $\mathcal{O}(\theta^2)$, i.e. with the result obtained by summing the partial widths of $\mathcal{B}_0$ decaying through mixing with $V_0$ (with $\Gamma_{V,0}$ evaluated at $s = m_{\mathcal{B},0}^2$), which then decays into SM final states \cite{Kuhn:1985eu,Kuhn:1987ty}. Throughout this paper, we will therefore qualitatively discuss the nature of the $\mathcal{B}$ resonance using the perturbative picture, while taking the mixing fully into account quantitatively. We will also not make a distinction between $\mathcal{B}_0$ and $\mathcal{B}$ or $V_0$ and $V$, unless we are explicitly discussing the mixing. 

\subsection{Collider Signatures} 
\label{sec:Collider}

There are three important classes of collider signatures for models of the type we have discussed:
(i)~mono-$X$, where the DM state $\chi$ is produced and observed as MET recoiling against a SM final state such as $X=j,h,W,Z\,$;
(ii)~$V$ resonant production with decaying channels such as dilepton, dijet or any other SM final states, and
(iii)~$\cB$ resonant production with $m_\cB\approx 2m_\chi$, decaying into a pair of leptons or jets.

The three channels probe different physics, as well as different regions of the dark sector parameter space. The mono-$X$ channel is an unavoidable signature of DM. The properties of the $\mathcal{B}$ resonance are completely determined by the DM mass and its self-interaction through the light mediator $Y$; therefore, by analyzing its properties, we study the DM directly. The $V$ resonance on the other hand probes the structure of the dark sector, but is not directly related to the puzzle of DM.


The mono-$X$ signature has been discussed previously~\cite{Abdallah:2015ter,Abdallah:2014hon,Bai:2010hh,Goodman:2010yf,Fox:2011pm}, and there are on-going searches at the LHC. We will demonstrate that for the models we consider, mono-jet searches probe a different region of parameter space than bound state resonance searches.

The production rate for the bound state $\cB$ at a $pp$ collider is given by (see ~\cite{Tsai:2015ugz,Kats:2009bv,Petrelli:1997ge} and also Appendix~\ref{sec:BoundStateTheory})
\begin{align}
	\label{eqn:qqbartoB}
	\sigma_{\cB} 
&\approx 	\sum_{q} \zeta(3) \frac{8\pi^2 (2J+1)}{9m_{\mathcal{B}}^3}  \Gamma_{\mathcal{B} \to q \overline{q}} \mathcal{L}_{q\bar{q}}\left( \tau_\cB \right),
\end{align}
where $J$ is the spin of the bound state; for a bound state produced from a vector mediator, $J=1$. $\zeta(s)$ is the Riemann $\zeta$-function, which takes into account the cross section for the production of all of the excited states of the bound state. Here $\tau_\cB \equiv m^2_\cB/s$, $m_\cB$ is the mass of the bound state, and $\sqrt{s}$ is the collider center-of-mass energy. $\cL_{q\bar{q}}$ is the parton luminosity function defined as 
\begin{alignat}{1}
    \label{eqn:PDF}
    \mathcal{L}_{q \overline{q}}(\tau) = \tau  \int_\tau^1 \frac{dx}{x}  f_q(x) f_{\overline{q}} (\tau/x) \, ,
\end{alignat}
with $f_q(x)$ being the parton distribution functions~(PDF), taken from~\cite{Martin:2009iq} for calculations in this paper.

In the perturbative limit, we can write
\begin{align}
    \label{eqn:qqbartoBPerturb}
    \sigma_{\cB} 
&\approx  \sum_{q} \frac{8 \pi \zeta(3)}{3 m_\cB}  \frac{g_q^2 g_\chi^2 |\psi(0)|^2 \mathcal{L}_{q\bar{q}}\left(\tau_\cB \right)}{(m^2_\cB - m^2_V)^2 +\Gamma_V^2(s = m_\mathcal{B}^2) m_V^2}  \, ,
\end{align}
where $g_q\,(g_\chi)$ sets the coupling of the mediator $V$ to quarks\,(DM) and $\psi(0)$ is the wave function of the bound state at the origin. 
For a Coulomb-like potential with coupling $\alpha_\cB$ (i.e. where the mass of the bound state mediator $m_Y$ can be neglected), $|\psi(0)|^2 = \alpha_\cB^3 m_\chi^3/8\pi$. 

This perturbative $\cB$ production cross section can be understood in three limits:
(i)~the heavy mediator limit, $m_V\gg m_\cB$;
(ii)~the light mediator limit, $m_V\ll m_\cB$, and
(iii)~$m_V\approx m_\cB$. 
The cross section in each limit is
\begin{multline}
	\label{eqn:qqbartoBlimit}
	\sigma_{\cB} = \frac{4 \pi \zeta(3)}{3 m_\chi} g_\chi^2 |\psi(0)|^2  \\
	\times \sum_{q} g_q^2 \mathcal{L}_{q \overline{q}} \left(\tau_{\mathcal{B}} \right) \begin{cases}
		\frac{1}{m_V^4}, & m_V \gg m_{\cB}, \\
		\frac{1}{m_\cB^4}, & m_V \ll m_{\cB}, \\
		\frac{1}{\Gamma_V^2 m_\cB^2}, & m_V \approx m_{\cB}.
	\end{cases}
\end{multline}
These equations show that the $\cB$ production cross section is enhanced when its mass is close to the mediator mass, and suppressed in the other two limits. Thus, we expect stronger sensitivity in this channel when $m_V \approx m_\cB \,$. Moreover, if $m_\cB \gg m_V$, which is in the limit where $V$ can also support dark matter bound states, the $\cB$ production cross section is suppressed by $\Gamma_V^2/m^2_\cB$  relative to the $m_V\approx m_\cB$ region. We also can see that for models where $\mathcal{B}$ is heavy enough to decay primarily into two or three $V$'s which then decay into 4 or 6 leptons at the LHC, the production rate of the bound state is suppressed relative to the regime where $m_V \approx m_\cB$.  

The mediator production cross section, $V$, is
\begin{align}
	\sigma_V 
\approx& 	\sum_q  \frac{8\pi^2}{3}\frac{\Gamma_{V\to q\bar q}}{m^3_V} \cL_{q\bar q}\left( \tau_V \right),
\end{align}
where $\tau_V \equiv m^2_V/s$. 
As we pointed out above, the $V$ resonance search does not directly probe the dark matter content. Further searches must be used to uncover the dark sector after discovering the mediator between the SM and the dark sector. Most importantly, when $m_V > 2m_\chi$ and $g_\chi\gg g_q, g_\ell$~(the coupling to leptons), the branching ratio of $V$ to SM particles becomes small, and resonance searches for $V$ grow ineffective. The full mixing calculation also bears out this conclusion: once $m_{V,0} > 2 m_\chi$, the $V$ resonance is heavier and lies above the $\chi \overline{\chi}$ threshold and is a wide resonance, while the lighter $\mathcal{B}$ resonance remains narrow and below the threshold. 

The comparison between mono-$X$ and bound state production is more complicated as the backgrounds for the two searches are different, and a more detailed comparison is required; we will show results for some specific models below.  
On generic grounds, the mono-$X$ cross section is reduced because of the PDF price of the additional jet. However, the two production cross sections scale as $\alpha_s g^2_q g^2_\chi$ and $\alpha_\cB^3 g^2_q g^2_\chi$, for the mono-$X$ and bound state cases respectively. Thus for $\alpha^3_\cB \ll \alpha_s$ we expect a reduced sensitivity  in the bound-state searches; this suggests $\alpha_\cB$ rather close to 1 will be required to make bound-state searches competitive. Moreover, the mono-jet search becomes ineffective once $m_V < 2m_\chi$, since the mono-jet process must then proceed through an off-shell $V$. 

In summary, the mono-jet search probes the region of parameter space where $m_V > 2 m_\chi$, while the $V$ resonance search is more sensitive to the region where $m_V < 2 m_\chi$. The bound-state production cross section, on the other hand, is enhanced precisely in the intermediate region, and outperforms the other two searches when $m_V \gtrsim 2 m_\chi$. These three searches are thus complementary, and probe different parts of parameter space, as we will show explicitly in our models below.

\subsection{Direct Detection Limits}
\label{sec:DDL}

Direct detection searches are very sensitive probes of DM, especially for DM with substantial couplings to hadrons, and mass at the EW scale or higher. 
Thus, viable models of dark resonance signals at the LHC must evade direct detection bounds. 

A naive estimate of the DM-nucleon scattering cross section at tree level, in terms of the parameters discussed in the previous subsection, gives $\sigma \sim g_q^2 g_\chi^2 m_N^2 /m_V^4 \sim 10^{-40} \text{cm}^2 g_q^2 g_\chi^2 \left( \text{TeV}/m_V\right)^4 $,  assuming $m_V$ is much larger than the typical momentum transfer in the scattering, and $m_\chi$ is much larger than the nucleon mass $m_N$. 
For comparison, under standard assumptions, the limit from XENON\,1T on this scattering cross section is of order $10^{-45} \text{cm}^2 (m_\chi/\text{TeV})$ \cite{Aprile:2017iyp}. 
Thus, if the elastic scattering spin-independent cross section is unsuppressed, we infer that the product of couplings $g_q^2 g_\chi^2 \lesssim 10^{-5} m_V^4 m_\chi/\text{TeV}^5$. 
This simple estimate is broadly consistent with more carefully obtained limits on a dark sector interacting with nucleons through a vector mediator for current and future direct detection experiments~\cite{DEramo:2016gos,Buchmueller:2014yoa}. 
Reasonably large couplings and sufficiently low dark sector masses are necessary for the significant production of the bound state resonance, but this parameter region of interest ($g_q g_\chi \sim 1$ and $m_V \sim 2 m_\chi \sim 1-4$ TeV) is generically in tension with direct detection bounds.

\begin{table}
    \begin{center}
    \begin{tabular}{cc}
        \begin{tikzpicture}
          \begin{feynman}
            \vertex (a1) {\(\chi_1\)};
            \vertex[right=1.5cm of a1] (a2); 
            \vertex[above=0.1cm of a2] (labelalphaD) {\(g_\chi\)};
            \vertex[right=1.3cm of a2] (a3) {\(\chi_{2}\)};
            \vertex[crossed dot, scale=1, below=1.3cm of a2] (b2) {\( \)};
            \vertex[below=0.3cm of b2] (labeleps) {\(g_q\)};
            \vertex[left=1.5cm of b2] (b1) {\(q\)};
            \vertex[right=1.5cm of b2] (b3) {\(q\)};
            \diagram*{
                (a1) -- [fermion] (a2);
                (a2) -- [fermion] (a3);
                (a2) -- [boson, edge label'=\(V\)] (b2);
                (b1) -- [fermion] (b2);
                (b2) -- [fermion] (b3);
            };
          \end{feynman}
        \end{tikzpicture} &
        \begin{tikzpicture}
          \begin{feynman}
            \vertex (a1) {\(\chi_1\)};
            \vertex[right=1.1cm of a1] (a2); 
            \vertex[right=1.5cm of a2] (a3);
            \vertex[above right=0.1cm and 0.5cm of a2] (labelchi2) {\(\chi_2\)};
            \vertex[right=0.9 of a3] (a4) {\(\chi_{1}\)};
            \vertex[crossed dot, scale=1, below=1.3cm of a2] (b2) {\( \)};
            \vertex[crossed dot, scale=1, below=1.3cm of a3] (b3) {\( \)};
            \vertex[left=1.05cm of b2] (b1) {\(q\)};
            \vertex[right=1.05cm of b3] (b4) {\(q\)};
            \vertex[below=0.4cm of b2] (buffer) {\( \)};
            \diagram*{
                (a1) -- [fermion] (a2);
                (a2) -- [fermion] (a3);
                (a3) -- [fermion] (a4);
                (a2) -- [boson] (b2);
                (a3) -- [boson] (b3);
                (b1) -- [fermion] (b2);
                (b2) -- [fermion] (b3);
                (b3) -- [fermion] (b4);
            };
          \end{feynman}
        \end{tikzpicture}
    \end{tabular}
    \end{center}
    
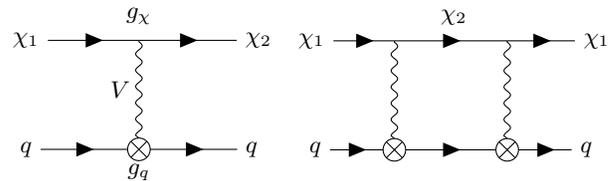
\captionof{figure}{Direct detection Feynman diagrams for inelastic DM models, with (left) tree-level inelastic scattering, and (right) one-loop, elastic scattering off nuclei in these experiments.}
    \label{fig:feynmanDirectDet}
\end{table}

However, any suppression to the naive tree-level cross section can alleviate this tension. As mentioned above, a simple scenario (``inelastic dark matter'') that leads to suppressed direct detection signals posits that the coupling between the DM $\chi_1$ and the mediator $V$ involves an unstable partner particle $\chi_2$, and the mass splitting between the DM and its partner is greater than the maximum kinetic energy of DM particles in the halo~\cite{Smith:2001hy,TuckerSmith:2004jv}. 
Fig.~\ref{fig:feynmanDirectDet} shows the relevant Feynman diagrams for direct detection of the DM particles. 
Such models also have interesting consequences for bound state formation at the LHC: if the bound state is produced in the $s$-channel from the mediator $V$, it will automatically be composed of the DM and its partner particle, or may only involve dark sector particles in the same multiplet as the DM, and not the DM at all. 

In such models, elastic scattering can still occur, but only at loop level. The direct-detection spin-independent cross section for scattering off a nucleon with target mass $m_T$ is given by~\cite{Kurylov:2003ra}
\begin{align}
	\sigma_{\rm SI} 
=	 \frac{4}{\pi} \left( \frac{m_\chi m_T}{m_\chi + m_T} \right)^2 (n_p f_p + n_n f_n)^2 \, ,
\end{align}
where $n_{p,n}$ are the number of protons and neutrons respectively, and $f_{p,n}$ are the corresponding matrix elements. 
We can generalize the effective operator analysis of~\cite{Hisano:2010fy}, to the dark sector models that will be of interest to us. Then 
\begin{align}
	\frac{f_N}{m_N} 
=	\sum_{q} \left(  f_q f_{T_q} + \frac{3}{4} (q(2) + \bar{q}(2)) (g_q^{(1)} + g_q^{(2)})  \right) \, ,
\end{align} 
where $N = n,p$, and we sum over $u,d,s$ quarks. Here the first term comes from a 1-loop diagram involving the Higgs, while the second term is a box diagram with two $V$ propagators. 
$q(2)$ and $\bar{q}(2)$ are the second moments of the quark and anti-quarks PDFs and $f_{T_q} = \langle N | m_q \bar{q}q | N\rangle / m_N$ is the nuclear form factor. For these we use the numeric values from~\cite{Hisano:2010fy}.
In the wino scenario discussed in \cite{Hisano:2010fy} these two contributions $f_q$ and $g_q^{1,2}$ are non-negligible but of opposite sign, thus leading to a cancellation. 
In the case of the dark sector models that we will consider later, the second contribution is suppressed by the small coupling between the SM mediator and the SM, while the first contribution will be negligible as the dark sector coupling to the SM Higgs will always be very small. 
Explicitly, we can write $g_q^{(1,2)} = (g_\chi^2 e \epsilon c_W Q / 4 \pi m_V^3) g_{T(1,2)} (m_V^2/m_\chi^2)$, where $g_{T(1,2)}$ are functions computed in~\cite{Hisano:2010fy}, $\epsilon$ is the small mixing parameter, and $Q$ is the charge of the quark. We find that in regions of parameter space of interest to the present work the contribution from loops to direct detection is thus no larger than $\sigma_{\rm SI} \sim10^{-48} \text{ cm}^2$ and is thus unconstrained.

\subsection{Overclosure and Indirect Searches}
\label{sec:OverInD}

In general, if annihilation to the light mediators that support the bound state is allowed, this process will tend to dominate freezeout. The same attractive potential that permits bound state formation will also generically enhance annihilation through the Sommerfeld enhancement~\cite{Hisano:2003ec,Hisano:2004ds}, potentially giving rise to large indirect signals in the present day. 
Formation of bound states followed by their decay can also significantly enhance indirect signals (\textit{e.g.}~\cite{Pospelov:2008jd}), if the mediator supporting the bound state is light enough that radiative capture of two DM particles into the bound state is kinematically allowed.

Let us first note that there are several possible annihilation channels which are $p$-wave suppressed at late times~\cite{Kumar:2013iva}; if these processes dominate freezeout, the late-time indirect detection signals will generally be very suppressed. We will see an example of this when we consider a model where the dominant annihilation is of Majorana fermions to light scalars. Furthermore, if DM-DM scatterings experience a repulsive potential rather than an attractive one, the DM-DM annihilation will be exponentially suppressed at low velocities~\cite{Cirelli:2007xd}. Note that in order for bound-state searches to be interesting in such a scenario, there must be at least one other particle with which the DM can form a bound state, and the DM must have an attractive interaction with that particle. 
In this initial discussion, therefore, we assume the dominant annihilation is $s$-wave and experiences an attractive Sommerfeld enhancement, to explore when indirect searches can set interesting constraints.

Let us consider the simple case where the effective potential experienced by the DM is a Yukawa potential, as discussed above. The $s$-wave Sommerfeld enhancement for a Yukawa potential with coupling $\alpha_\cB$ and mediator mass $m_Y$ can be well approximated by~\cite{Cassel:2009wt, Slatyer:2009vg}:
\begin{alignat}{2}
	S \,
&= \,\,&& 	\frac{2 \pi \alpha_\cB}{v_\text{rel}} 
	\frac{\sinh\left(\frac{6}{\pi} \frac{m_\chi v_\text{rel}}{m_Y}\right)}{\cosh\left(\frac{6}{\pi} \frac{m_\chi v_\text{rel}}{m_Y}\right) - \cos\theta} \nonumber\\
&\ge \,\, && 	\frac{2 \pi \alpha_B}{v_\text{rel}} 
	\frac{\sinh\left(\frac{6}{\pi} \frac{m_\chi v_\text{rel}}{m_Y}\right)}{\cosh\left(\frac{6}{\pi} \frac{m_\chi v_\text{rel}}{m_Y}\right) +1} \, , 
	\label{eq:sommerfeld} 
\end{alignat}
where $v_\text{rel}$ is the relative velocity between the DM particles, and in the second line we use $\cos\theta> -1$ with $\theta = 2\pi\sqrt{\frac{6}{\pi^2} \frac{m_Y}{\alpha_\cB m_\chi} - \left(\frac{3 m_\chi v_\text{rel}}{\pi^2 m_Y}\right)^2}$ an angle controlling the resonance positions. The inequality is saturated for real values of $\theta$, at the minima between resonances where $\cos\theta = -1$. (It is also approximately saturated where $m_Y \rightarrow 0$ and $v_\text{rel} \ll \alpha_\cB$, where $S \approx   2 \pi \alpha_\cB/v_\text{rel}$.) 

Note that for fixed $m_\chi$, requiring the correct relic density fixes $\alpha_\cB$, if the assumptions are made that 
(a)~this channel dominates during freezeout, and 
(b)~the mass of the mediator is irrelevant during freezeout. 
The latter assumption is approximately true away from resonances and if $m_V/m_\chi$ is smaller than the typical velocity of particles around freezeout ($v\sim 1/3$). 
For large values of $m_Y$ (requiring large $\alpha_\cB$, since $m_Y/m_\chi < \alpha_\cB/1.68$), or values of $m_Y$ corresponding to resonant Sommerfeld enhancement ($\cos \theta \rightarrow 1$ as $v_\text{rel} \rightarrow 0$), freezeout may be more complicated and needs to be studied more carefully; we will include the full $m_Y$ dependence when we examine specific models.

However, if we consider $\alpha_\cB$ to be fixed given $m_\chi$, and hold $v_\text{rel}$ fixed, then our expression for the lower bound on the Sommerfeld enhancement is a monotonically decreasing function of $m_Y$; thus indirect detection will set a \textit{lower} bound on $m_Y$ (all values of $m_Y$ below this threshold will be ruled out). 
Since the requirement for bound state formation sets an \textit{upper} bound on $m_Y$, one can ask whether these two criteria are in conflict. 
Equivalently, requiring $m_Y < \alpha_\cB m_\chi/1.68$ implies that $S > \frac{2 \pi \alpha_\cB}{v_\text{rel}}\sinh\left(\frac{3.21 v_\text{rel}}{\alpha_\cB}\right)\Big/\left( \cosh\left(\frac{3.21 v_\text{rel}}{\alpha_\cB}\right) +1\right)$. 
In order for the model to avoid exclusion by indirect detection (except possibly where $m_Y$ is important to freezeout), this minimal Sommerfeld enhancement must be permitted by the data. Note that for $\alpha_\cB \gg v_\text{rel}$ ( $v_\text{rel} \sim 10^{-3}$ in the present-day Milky Way halo), this minimal enhancement will reduce to an $\alpha_\cB$- and $v_\text{rel}$-independent prefactor of:
\begin{align} 
	S_\text{min} = 2\pi \times 3.21/2 \approx 10.
\end{align}
This minimal Sommerfeld-enhanced cross section is rather close to indirect detection bounds for a DM species that comprises 100\% of the DM and whose abundance is set by thermal freezeout, for DM masses below $\sim 1$ TeV ({\it e.g.} \cite{Fermi-LAT:2016uux, Abdallah:2016ygi}); permitting $m_Y \lesssim m_\chi v_\text{rel}$ would generally significantly overproduce limits from indirect detection, unless $Y$ decays primarily into invisible channels. If we assume $m_Y \gg m_\chi v_\text{rel}$ in the present day, then we can approximate $S \gtrsim 6 \alpha_\cB m_\chi/m_Y$, and thus if the maximum allowed Sommerfeld factor is $S_\text{max}$, then $m_Y \gtrsim 6 \alpha_\cB m_\chi / S_\text{max}$. Of course, smaller values for $m_Y$ are permissible if the species that forms bound states comprises only a small fraction of the overall dark matter density.

If the dominant annihilation channel consists of $s$-wave annihilation to mediators coupled to the DM with strength $\alpha_\mathcal{B}$, then the annihilation cross section at low velocities is of order $\langle \sigma v_\text{rel} \rangle \approx \pi \alpha_\mathcal{B}^2/m_\chi^2$ (this expression is exact for Dirac or pseudo-Dirac DM annihilating to U(1) dark gauge bosons). Requiring that this cross section fall below the thermal value of $\langle \sigma v_\text{rel} \rangle \approx 2\times10^{-26}$ cm$^3$/s $\approx 1.7 \times 10^{-9}$ GeV$^{-2}$ suggests an overclosure bound of $m_\chi \lesssim \alpha_\mathcal{B} \times 43 \, \text{TeV}$. As we will see, we will generally be interested in masses around a few TeV and $\alpha_\mathcal{B} \gtrsim 0.1$, so the overclosure bound will not typically be particularly constraining. 
This estimate ignores Sommerfeld enhancement and bound state formation during freezeout, which can be important \cite{Feng:2010zp,vonHarling:2014kha}. For $\alpha_B \gtrsim 0.1$, the Sommerfeld enhancement is non-negligible during the freezeout epoch; however, for attractive Sommerfeld enhancement, including this effect only reduces the late-time relic abundance. This further relaxes the overclosure bound, and since it reduces the abundance of the species in question, also weakens constraints from indirect detection. (However, it makes it more challenging to generate 100\% of the DM abundance by the same species that forms bound states.)

Likewise, radiative formation of bound states can also contribute to the depletion of DM at early times and indirect signals at late times \cite{Pospelov:2008jd, MarchRussell:2008tu,vonHarling:2014kha,An:2016kie}. These radiative processes are only kinematically unsuppressed if enough energy is available to produce an on-shell light mediator, i.e. the binding energy + kinetic energy of the particles is greater than $m_Y$. Bound state formation can also occur through radiation of an off-shell heavy mediator that decays to SM particles, but such processes will be suppressed by a small mixing with the SM and also by the mass of the heavy mediator. 
Thus, there are two distinct regimes for $m_Y$ from an indirect-detection perspective: $\alpha_\mathcal{B} m_\chi/1.68 \gtrsim m_Y \gtrsim \alpha_\mathcal{B}^2 m_\chi/4$, where bound states exist but radiative capture into them is suppressed, and $m_Y \lesssim \alpha_\mathcal{B}^2 m_\chi/4$, where radiative capture processes are unsuppressed. We will ignore bound-state effects in the former case, but account for their impact on indirect-detection signatures in the latter case.

However, we will ignore the effects of bound-state formation during freezeout. A careful treatment of bound-state effects during freezeout requires accounting for dissociation of the bound states through interactions with the light-mediator bath. If $m_Y \lesssim \alpha_\mathcal{B}^2 m_\chi/4$, then for $\alpha_\mathcal{B} \lesssim 0.5$ we expect the temperature at freezeout to be comparable to or larger than the binding energy (taking the standard estimate $T_\text{freezeout} \sim m_\chi/20$), and so dissociation effects could be substantial. Thus while the presence of radiative capture into bound states during freezeout may further deplete the DM abundance, relaxing both the overclosure and indirect limits further, a full calculation would require a careful analysis (as performed in \textit{e.g.} Ref.~\cite{vonHarling:2014kha}).

We will show that the indirect detection constraints and overclosure limit cannot fully exclude the regions of parameter space relevant to collider searches for the bound states, even without taking the impact of bound-state effects on freezeout into account, for both models we consider. Since including the bound-state effects during freezeout would only relax these constraints further, we are justified in neglecting them for purposes of this work.

\subsection{Dark Matter Self-Interactions}
\label{sec:DMSI}

Constraints on DM self-interactions require that $\sigma/m_\chi \lesssim 1$ cm$^2$/g $\approx 1/(60 \text{MeV})^3$~\cite{Robertson:2016xjh, Tulin:2017ara, Harvey:2015hha}. 
As we are interested in the regime where a long-range potential exists and can support bound states, we cannot use the Born approximation to estimate scattering rates. However, if $m_Y/m_\chi \gtrsim v_\text{rel}$ while still satisfying Eq.~\eqref{eqn:boundStateRequirement}, the typical relative velocity of DM particles in galaxies and galaxy clusters, then we can make the approximation that $s$-wave scattering dominates and use the analytic estimate for the scattering cross section derived in Refs.~\cite{Tulin:2013teo,Schutz:2014nka}.

The scattering cross section is approximated in the low-velocity limit by~\cite{Tulin:2013teo,Schutz:2014nka}
\begin{align} 
	\sigma_T 
&= 	\frac{4\pi}{(m_\chi v_\text{rel})^2} \left|1 - e^{2 i \delta} \right|^2 \, , 
\end{align}
where $\delta = -\left[2 \gamma + \ln(c) + \pi \cot(\pi\sqrt{c}) \right] a c$,  $a = v_\text{rel} / 2 \alpha_{\cB}$, $c  = \alpha_{\cB} m_\chi/1.6 m_Y$ and $\gamma \approx 0.577$ is the Euler-Mascheroni constant.
We see that away from resonances, which occur when $\cot(\pi \sqrt{c})$ diverges, the size of the phase shift is controlled by $a c = m_\chi (v_\text{rel}/2)/(1.6 m_Y)$. The regime where the $s$-wave contribution dominates is thus a regime where (away from resonances) this phase shift is small, and we can write:
\begin{align} \sigma_T \sim \frac{4 \pi}{(m_\chi v_\text{rel}/2)^2} a^2 c^2 \sim \frac{4 \pi}{ m_Y^2}, \end{align}
which is just the geometric cross section.

Assuming this geometric cross section, we see that the self-interaction bound will be satisfied provided $(m_\chi m_Y^2)^{1/3} \gtrsim 100$ MeV, which for 1 TeV DM requires only that $m_Y \gtrsim 1$ MeV. 
Thus, away from points in the parameter space where there is a near-zero-energy bound state, we expect the self-interaction rate to be undetectable, despite the rather large couplings we invoke.

\section{Dark Sector Models With Detectable Bound States}
\label{sec:darksector}

We now consider two examples of phenomenologically viable DM models containing bound states, which can lead to interesting signatures at the LHC. These models serve as examples of how to build non-supersymmetric models that realize the requirements laid out in Sec.~\ref{sec:PhenoBound}. 
We will show that in these models, the search for bound-state resonances can probe parameter space which is not accessible to mono-jet searches and resonance searches for the mediator.

In the first model, which we label as the ``pseudo-Dirac'' model, the dark sector consists of a pair of almost-degenerate Weyl fermions that are charged under a dark-sector U(1)$_D$ gauge group, which is broken by a dark Higgs-like scalar. These fermions can form bound states with the dark Higgs as the mediator. 
The second model, which we refer to as the ``triple Higgs'' model, is based on a completely broken SU(3)$_D$ gauge theory, with the dark matter candidate being a Dirac fermion in the fundamental of the gauge group. Much of the phenomenology of this model, including bound state formation and couplings to the SM, is derived from the symmetry breaking pattern of the theory, with both the mediator that supports the bound state and the mediator to the SM being massive gauge bosons of the broken SU(3)$_D$ group. 
In both cases, the dark sector interacts with the SM via a vector portal with kinetic mixing, and the DM direct detection cross section is suppressed by the fact that at tree-level the DM scatters into a heavier state.

Before introducing these models, however, we will consider a simpler scenario that is familiar from SUSY, that of pure wino/higgsino DM (Sec.~\ref{sec:winoHiggsino}). We will show that the production rate of wino/higgsino-onium bound states at the LHC is too small to be constraining, but this scenario shares many of the properties of our more complicated dark-sector models and thus has pedagogical value. We will then review the details of kinetic mixing between new dark gauge bosons and the SM neutral gauge bosons (Sec.~\ref{sec:KineticMixing}), since this mechanism describes the leading interaction of the SM with the dark sector in both dark-sector scenarios we consider, before describing in detail the two models (Sec.~\ref{sec:PseudoDirac} and~\ref{sec:TripleHiggs}).


\subsection{A Weakly Interacting Example: SU(2)$_L$ Minimal Dark Matter}
\label{sec:winoHiggsino}

Sub-TeV superpartners of the EW bosons and of the two Higgs doublets in SUSY theories can potentially be produced and detected at the LHC, with the lightest neutralino being a particularly well-motivated, weakly-interacting DM candidate. Outside of SUSY theories, models of ``minimal dark matter'' where the DM transforms under a low-dimensional representation of SU(2)$_L$ have similar phenomenology to neutralino DM~\cite{Cirelli:2005uq, Cirelli:2007xd, Fukuda:2017jmk}. Pure wino or higgsino DM corresponds to the lowest-lying mass eigenstates from, respectively, an SU(2)$_L$ triplet of Majorana fermions or an SU(2)$_L$ Dirac fermion doublet with hypercharge $1/2$. The hypercharge-zero SU(2)$_L$ quintuplet is also a viable ``minimal dark matter'' candidate.

If the DM transforms as part of a SU(2)$_L$ multiplet, then it will be accompanied by heavier charged partner particles in the same multiplet. After EW symmetry breaking, the wino triplet separates into a lighter neutral Majorana fermion $\chi^0$ and a heavier charged Dirac fermion $\chi^\pm$; the higgsino multiplet gives rise to two neutral Majorana states $\chi^1$, $\chi^2$ and a charged Dirac fermion $\chi^\pm$. These charged partners can always form Coulombic bound states; when the DM is sufficiently heavy, $W$ and $Z$ exchange may also support bound states including the DM itself (\textit{e.g.}~\cite{Asadi:2016ybp}). Numerical calculations indicate that for wino DM there is a crossover point at a DM mass of around 5\,TeV, where the ground state transitions from being primarily composed of $\chi^+ \chi^-$ bound by photon exchange, to being composed of an admixture of $\chi^0 \chi^0$ and $\chi^+ \chi^-$ bound by the gauge bosons of an approximately unbroken SU(2)$_L$ symmetry.

SU(2)$_L$-DM models have many attractive features of the type discussed in Sec.~\ref{sec:PhenoBound}, and behave as prototypes for the models of interest to us. They naturally possess multiple mediators, one of which is massless and supports bound states, while the other massive mediators are all known particles in the SM. The SU(2)$_L$ multiplets contain several states nearly-degenerate with the DM; the couplings of the gauge bosons with the DM and its partners are naturally off-diagonal, and so the elastic scattering relevant to direct detection proceeds only at one-loop level (and also suffers from additional cancellations which suppress the rate further \cite{Hill:2013hoa}). Direct and indirect constraints on wino and higgsino DM have been studied extensively; thermal wino DM constituting 100\% of the DM is in tension with H.E.S.S. observations of the Galactic Center ({\it e.g.}~\cite{Cohen:2013ama, Fan:2013faa,Hryczuk:2014hpa,Ovanesyan:2016vkk,Cuoco:2017iax,Abramowski:2013ax,Baumgart:2017nsr}), but a subdominant wino DM contribution at lower wino masses is difficult to exclude. 
Pure higgsino DM is not currently experimentally testable by either direct or indirect detection~\cite{Krall:2017xij,Baumgart:2015bpa}.

A complicating factor in SU(2)$_L$ DM models is the presence of multiple mediators that can potentially support a bound state, which become most important if the DM is heavy enough that $\alpha_W m_\chi \gtrsim m_W, m_Z$ with $\alpha_W=g_W^2/4\pi\approx1/30$. 
In this case, there is a long-range potential that mixes the two-body DM--DM state with other particle anti-particle states, i.e. $\chi^0 \chi^0$ mixes with $\chi^+ \chi^-$ in the wino case, and $\chi^1 \chi^1$ can mix with $\chi^2 \chi^2$ and $\chi^+ \chi^-$ in the higgsino case. 
This can lead to $\chi^+ \chi^-$ states that are only pseudo-bound, despite the presence of the photon-mediated Coulomb potential: if the combined $W/Z/\gamma$-exchange potential is not deep enough to also bind the $\chi^0 \chi^0$ component, then the $\chi^+ \chi^-$ state (or {\it e.g.} $\chi^{++} \chi^{--}$ in representations, such as the quintuplet, where higher-charge states exist) may decay rapidly to unbound $\chi^0 \chi^0$ through $t$-channel exchange of $W$ bosons. Parametrically, the cross section for $\chi^+ \chi^- \rightarrow \chi^0 \chi^0$ through this channel, for heavy DM with $m_\chi \gg m_W$, is $\sigma v_\text{rel} \sim \sqrt{\Delta/m_\chi} \alpha_W^2 m_\chi^2/m_W^4$, where $\Delta$ is the available energy (i.e. the splitting between the mass of the $\chi^+ \chi^-$ two-body state, including any binding energy, and the mass of the final $\chi^0 \chi^0$ state). By comparison, the cross section for annihilation to SM quarks, leptons and gauge bosons is of order $\sigma v_\text{rel} \sim \alpha_W^2/m_\chi^2$. Thus we expect the former to dominate over the latter when $\sqrt{\Delta/m_\chi} \gtrsim (m_W/m_\chi)^4$.

However, there is an important caveat to this argument: in fermionic models of this type, this mixing between $\chi^0 \chi^0$ and $\chi^+ \chi^-$ occurs only in the states with even $L+S$ (where $L$ and $S$ are the quantum numbers describing the total orbital angular momentum and total spin of the bound state); states with odd $L+S$ have symmetric wavefunctions and cannot support two identical fermions. Since the mediator to the DM is an EW gauge boson, the bound state dominantly produced at colliders has $L=0$ and $S=1$; these are true bound states, not pseudo-bound, and cannot decay rapidly to pairs of identical DM particles. In particular, in the pure wino case the $\chi^+ \chi^-$ bound state with $L=0$, $S=1$, denoted $\mathcal{B}_w$, decays dominantly via an $s$-channel $\gamma/Z$ to SM fermion pairs or through a $t$-channel exchange of a $\chi^0$ into a $W^+ W^-$ final state~\cite{Asadi:2016ybp}; final states involving the DM are suppressed. We will see this behavior arise again in our example dark-sector models.

The pure higgsino limit serves as an example of a model where there are two neutral mass eigenstates that can be close in mass, denoted as $\chi^0_1$ and $\chi^0_2$, the lighter of which ($\chi^0_1$) is the DM. 
In this case, the decay of $\chi^+ \chi^-$ to $\chi^0_1 \chi^0_2$ may be allowed. If~$\Delta_{+0} \equiv 2m_{\chi^\pm} - m_{\chi_1^0} - m_{\chi_2^0} < 0$, the $\chi^+ \chi^-$ bound state never mixes into the $\chi_1^0 \chi_2^0$ from kinematic considerations. When $\Delta_{+0} > 0$ however, the $\chi^+ \chi^-$ can simply decay into free $\chi_1^0 \chi_2^0$, and if the width for this decay is significantly larger than the width of the $\chi^+ \chi^-$ bound state, the bound state is effectively never formed.\footnote{When the widths are comparable, bound state decays into $\chi_1^0 \chi_2^0$ becomes an additional decay channel, together with decays to $W^+W^-$ or SM fermions.}
Thus, for the pure higgsino case the sign of the parameter $\Delta_{+0}$ is critical to the bound state phenomenology, at least for DM masses below the TeV scale. 
This parameter is positive when the lightest neutralino is a pure higgsino and both the wino and the bino are taken to be infinitely massive, but there exists a range of SUSY-breaking parameters which can produce a lightest neutralino that is almost purely higgsino with a significantly more massive bino and wino, while having $\Delta_{+0} < 0$~\cite{Han:2014kaa}. 
With this choice, a $\chi^+ \chi^-$ higgsino bound state $\mathcal{B}_h$ can be formed and can decay in the same way as $\mathcal{B}_w$, albeit with different coupling constants to the EW bosons. 

\begin{figure}
    \centering
    \includegraphics[scale=0.58]{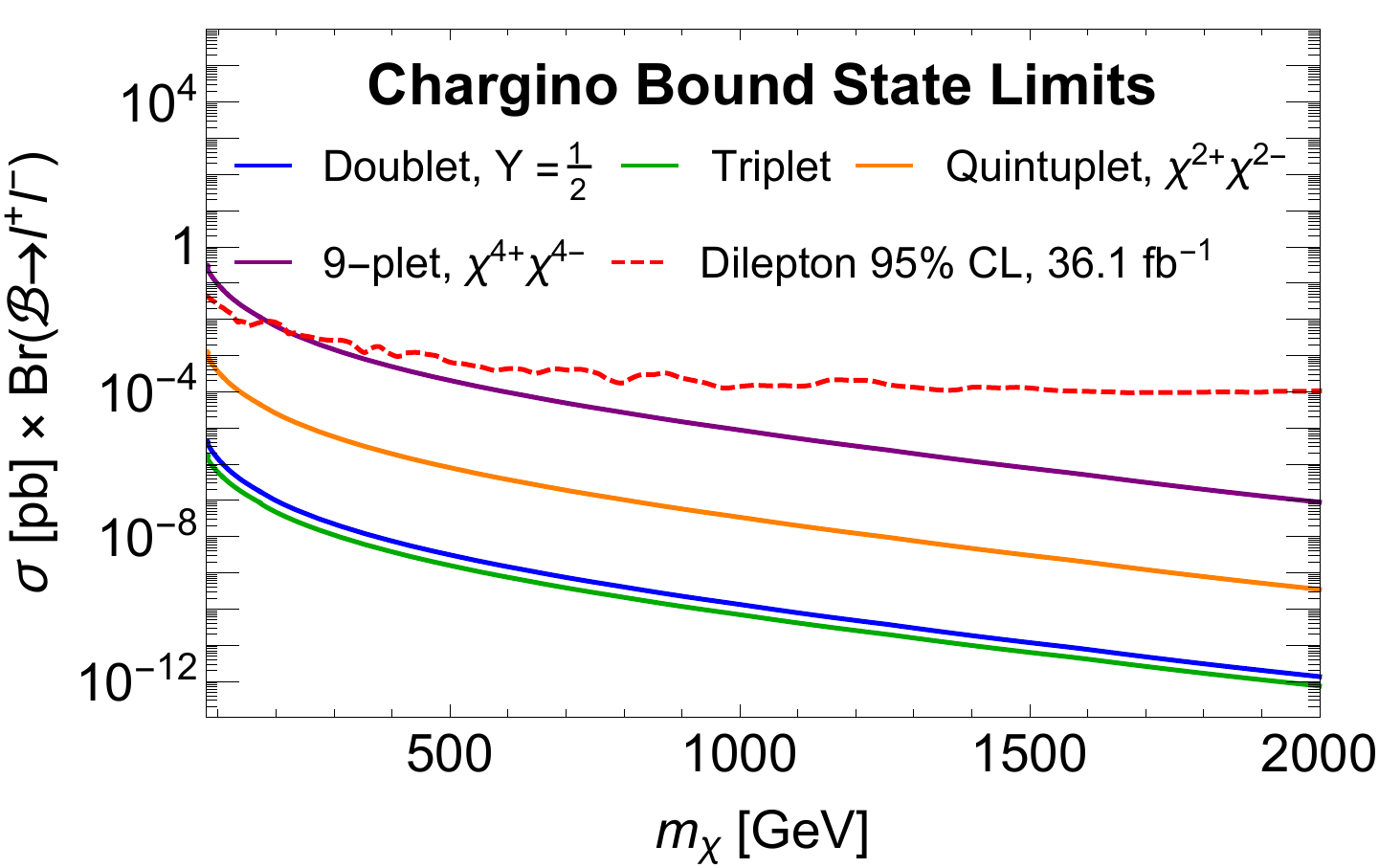}
    \caption{The production cross section times branching ratio into leptons for chargino-onium states made up of: (green) pure higgsino $\chi^+ \chi^-$; (blue) pure wino $\chi^+ \chi^-$; (orange) SU(2)$_L$ quintuplet, zero hypercharge $\chi^{2+} \chi^{2-}$, and (purple) SU(2)$_L$ 9-plet, zero hypercharge $\chi^{4+} \chi^{4-}$. The 95\% confidence limits from a dilepton resonance search for $Z'$ with \SI{36}{fb^{-1}} of data are also shown (red, dashed).}
    \label{fig:charginoonium}
\end{figure}

Unfortunately, if the DM is part of an SU(2)$_L$ doublet or triplet, the bound state production rate at the LHC is too small to be observed. This is due to the smallness of the EW couplings, which controls the production rate. Figure~\ref{fig:charginoonium} shows the production cross section times branching ratio into leptons of chargino-onium states for fermions charged under the EW gauge group in different representations. Chargino-onia from both pure winos and pure higgsinos have production cross sections that are far too small for dilepton searches at the LHC to be effective. However, for DM in a larger representation of SU(2)$_L$, fermions having large electromagnetic charges $Q$ can be produced. The production cross section of these states scales rapidly with $Q$, while the partial widths into SM particles remain unchanged. The enhancement factor relative to the pure wino is $Q^8$, with $Q^6$ coming from the wavefunction of the bound state at the origin $|\psi(0)|^2$, and an additional $Q^2$ from the coupling of these fermions to the $\gamma$ and $Z$. For charginos with $Q = 4$ in an SU(2)$_L$ 9-plet, the production cross section for the $\chi^{4+} \chi^{4-}$ chargino-onium becomes large enough to be probed by the current dilepton resonance search results. Such large representations are generally disfavored since they lead to non-perturbative values of $\alpha_W$ below the Planck scale~\cite{Cirelli:2005uq}; however, these results more broadly demonstrate that models with large coupling constants or large charges are particularly suited for bound state searches at the LHC. Searches for multi-charged lepton bound states decaying into two photons, for example, have been shown to be effective in searches for leptons with a sufficiently large hypercharge~\cite{Barrie:2017eyd}. 
%

\subsection{Kinetic Mixing}
\label{sec:KineticMixing}

We now turn our attention the dark sector models that we briefly described above. Both of the models we will consider interact with the SM through a vector portal, with kinetic mixing with the SM U(1)$_Y$:
\begin{multline}
	\mathcal{L}_{\text{kin-mix}} 
= 	-\frac{1}{4} V_{\mu\nu} V^{\mu\nu} - \frac{\epsilon}{2} B_{\mu\nu} V^{\mu\nu} - \frac{1}{4}B_{\mu\nu} B^{\mu\nu} \\
	 + \frac{1}{2} m_{V}^2 V_\mu V^\mu + \frac{1}{2} m_Z^2 Z_\mu Z^\mu \, ,
\end{multline}
where $V_{\mu\nu}$ ($B_{\mu\nu}$) is the field strength of the dark gauge boson (SM hypercharge), and we have included the mass term for both $V$ and the SM $Z$. Here, $V_{\mu\nu}$ can be non-abelian: such a mixing term appears in the triple Higgs model in the form of a dimension-5 operator $H^a_D V_{\mu\nu}^a B_{\mu\nu}$ where $H^a_D$ is an adjoint scalar that acquires a VEV, and $a = 1, \cdots, 8$ is an SU(3)$_D$ color index. 

This interaction can be diagonalized in the mass basis; a detailed description of this diagonalization procedure is discussed in~\cite{Cassel:2009pu,Hook:2010tw}, and reviewed in Appendix~\ref{app:kineticMixing}. In the non-abelian case, only the abelian portion of the field strength is diagonalized, with the non-abelian portion remaining as an interaction term in the model. The diagonalization introduces an $\epsilon$-suppressed coupling between the physical dark gauge boson and the SM electromagnetic, $J^\mu_{\text{EM}}$, and weak-neutral, $J^\mu_{Z}$, currents, as well as an $\epsilon$-suppressed coupling between the SM $Z$-boson and the dark sector current, $J^\mu_D$: 
\begin{alignat}{1}
	J^\mu_{\text{EM}} A_\mu &\to J^\mu_{\text{EM}} \left(A_\mu - \epsilon c_W V_\mu\right), \nonumber \\ 
	J^\mu_Z Z_\mu &\to J_Z^\mu \left(Z_\mu + \frac{\epsilon s_W }{1- r^2}V_\mu \right), \nonumber \\
	J^\mu_D V_\mu &\to J_D^\mu \left(V_\mu - r^2\frac{\epsilon s_W}{1 - r^2} Z_\mu \right),
    \label{eqn:currents}
\end{alignat}
where $A$ is the SM photon, $s_W$ ($c_W$) is the sine (cosine) of the weak mixing angle, and $r \equiv m_Z/m_{V}$. 
All of the fields are given in the mass basis: note that the DM fermionic current couples directly to the $Z$, so both $V$ and $Z$ mediate the production of dark sector particles with $q \overline{q}$ interactions, and both must be included in amplitude calculations.

The mixing between $V$ and $Z$ also shifts their masses by a fraction of $\mathcal{O}(\epsilon^2)$: the shift in the $Z$-mass has important consequences for EW precision constraints on these models which we will discuss below, but otherwise these shifts will be neglected for the rest of the paper. We will always assume that $r \ll1$ throughout in both models.

\subsection{U(1)$_D$ Pseudo-Dirac Dark Matter}
\label{sec:PseudoDirac}

We now consider a simple, viable dark matter model, where the bound state signature gives complementary information about the dark sector and probes different region of the parameter space than the mono-$X$ searches. 
Our model is based on the ``minimal model'' of ~\cite{Finkbeiner:2010sm} (loosely based on the ``excited dark matter'' scenario of ~\cite{Finkbeiner:2007kk}), but we use an ordinary Yukawa interaction between the dark Higgs and the fermions in the dark sector instead of a dimension-5 operator. 

This model contains a gauged U(1)$_D$ field, $V$, kinetically mixed with the SM U(1)$_Y$, a Dirac fermion $\Psi$ and a dark Higgs, which in unitary gauge can be written as $\Phi_D = (v_D + h_D)/\sqrt{2}$, with $v_D$ as its VEV. The U(1)$_D$ charges for the fermion $\Psi$ and $\Phi_D$ are 1 and 2 respectively. The Lagrangian is
\begin{alignat}{2}
	\cL_{\rm dark-Maj} 
&=&& \, i \overline{\Psi} \slashed{D} \Psi + (D_\mu \Phi_D)^\dagger(D^\mu \Phi_D)  - m_D \overline{\Psi}\Psi \nonumber\\
& &&	 - y_D \left( \overline{\Psi}^C \Psi \Phi_D^* + \text{h.c.} \right) 
	+\mathcal{L}_{\text{kin-mix}} \, ,
\label{eqn:pseudoDiracL}
\end{alignat}
where $D_\mu \equiv \partial_\mu - i g_D V_\mu$ is the covariant derivative for $\Psi$ and $D_\mu \equiv \partial_\mu - 2i g_D V_\mu$ is the covariant derivative for $\Phi_D$, with $C$ denoting charge conjugation. 
Following~\cite{Dreiner:2008tw}, we write $\Psi$ as a  Weyl fermion pair $(\chi, \eta^\dagger)$. Thus, the Yukawa interaction becomes
\begin{align}
	\cL_{Y_D}
=	-y_D \left(\chi\chi \Phi_D^* + \eta\eta \Phi_D +  \text{h.c.} \right) \, .
\end{align}
After the dark Higgs gets a VEV, the Yukawa interaction generates a fermion mass splitting. The fermion mass matrix is
\begin{align}
	\frac{1}{2}(\chi \ \eta)
	\begin{pmatrix}
		m_M & m_D \\
		m_D & m_M
	\end{pmatrix}
	\begin{pmatrix} \chi \\ \eta \end{pmatrix} + \text{h.c.}
\end{align}
with $m_M=\sqrt{2}y_D v_D$. The mass eigenstates are then given by $\chi_1 = (\eta + \chi)/\sqrt{2}$ and $\chi_2 = i(\eta - \chi)/\sqrt{2}$, with masses $m_{1,2} = m_M \pm m_D$. 
In the mass basis, the dark Yukawa interaction terms can be written as
\begin{align}
	\cL_{Y_D} = 
	- \frac{y_D}{\sqrt{2}}(v_D + h_D)\left(\chi_1 \chi_1 -\chi_2 \chi_2 + \text{h.c.} \right) \, ,
\end{align}
and the interaction with the dark photon is then given by
\begin{alignat}{1}
	-i g_D \left(V_\mu - r^2\frac{\epsilon s_W}{1 - r^2 } Z_\mu \right) \left(\chi_1^\dagger \overline{\sigma}^\mu \chi_2 - \chi_2^\dagger \overline{\sigma}^\mu \chi_1 \right) \, .
\end{alignat}
The interaction with the SM is thus off-diagonal, and the direct detection constraint is significantly relaxed because the $\chi_1$ - $\chi_2$ mass splitting means the elastic scattering cross section is suppressed at one-loop (and the one-loop contribution is expected to be small as previously discussed).

In this model, a DM bound state can be produced at the LHC through the process shown in Fig.~\ref{fig:feynmanMonoXAndResonance}, supported by the exchange of either a dark Higgs or a dark photon. We will focus on the case where the dark Higgs is light and supports the bound state, while the dark photon is heavier and is the principal mediator to the SM, in order to ensure a one-loop suppression in the direct detection cross section while maintaining a large coupling between the quarks and the mediator to the SM and a sizable branching ratio of the bound state to leptons. The dark Higgs is assumed to have some small mixing with the SM Higgs that allows it to decay.

\begin{figure}
    \centering
    \includegraphics[scale=0.3]{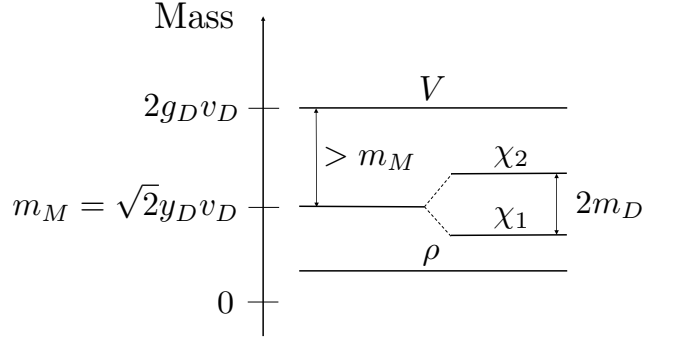}
    \caption{Spectrum of particles in the pseudo-Dirac model.}
    \label{fig:spectrum_PD}
\end{figure}

Because of the symmetry breaking pattern, there are only three independent parameters among~$\{m_\chi, m_V, \alpha_D, y_D\}$. The mass hierarchy required above can be achieved by choosing $m_D \ll m_M$, so that $m_D$ is the small mass splitting, and $m_{1,2} \simeq m_M$. The spectrum of particles in this model is shown in Fig.~\ref{fig:spectrum_PD}.

A large value of $\alpha_D$ leads to a Landau pole in a broken U(1) theory at a scale above $m_V$~\cite{Davoudiasl:2015hxa}. However, since we are mainly interested in the phenomenology of bound states below the scale $m_V$, we assume that a UV completion of the model will avoid the Landau pole. We will later discuss another model with a non-abelian gauge group in the dark sector which will avoid the need for a UV completion. 

If the dark bound state, $\mathcal{B}$, is produced from SM initial states, it must be produced from a $Z$ or $V$ exchange. 
Since the couplings of these gauge bosons to the dark Majorana fermions are off-diagonal, the resulting bound state must be composed of a $\chi_1$ and a $\chi_2$ particle, and for an $s$-wave state, it must have spin-1.
Moreover, since $h_D$ only couples $\chi_1$ to $\chi_1$ and $\chi_2$ to $\chi_2$, decays into final states containing only $h_D$ are forbidden, and if $m_V > m_\mathcal{B}$, the only available decay modes for $\mathcal{B}$ are through $V$ back into the SM particles. 

\subsection{SU(3)$_D$ Triple Higgs Model}
\label{sec:TripleHiggs}

We now consider a dark sector model based on a completely broken SU(3)$_D$ gauge theory, where all of the phenomenologically desirable properties of the dark sector emerge from the breaking pattern of the gauge symmetry. This model has some similarities with the non-abelian DM models of~\cite{Chen:2009ab}, featuring small mass splittings among the components of the DM multiplet that suppress the direct detection cross section. Because the mediator supporting the bound state is a vector in this model as opposed to a scalar in the pseudo-Dirac model above, the indirect detection constraints of the two models turn out to be quite different.

A completely broken SU(3) gauge group was chosen to allow for a sufficiently large gauge coupling, which is favorable for the production of bound states that are supported by gauge bosons.\footnote{DM models with an unbroken gauge group are constrained by the fact that dark matter is effectively collisionless in galactic dynamics~\cite{Ackerman:mha}.} A broken U(1) theory, such as the one found in the pseudo-Dirac model, with a coupling strength $\alpha_D \gtrsim 0.5$ at momentum scales above the gauge boson mass quickly runs into a Landau pole. Thus a broken U(1) theory with a large coupling constant is likely to have emerged from a larger, nonabelian gauge group in the first place~\cite{Davoudiasl:2015hxa}. 
We choose an SU(3) gauge group rather than SU(2), because for a fermion in the fundamental of a completely broken SU(2) theory with an off-diagonal coupling to the SM, the gauge boson corresponding to the diagonal generator produces a repulsive potential between the two components of the fermion, making it difficult for a phenomenologically viable bound state to exist without introducing additional light mediators. 

As in the previous model, the coupling between the dark sector and the SM is mediated by the mixing of the dark and SM gauge bosons; in this non-abelian case, the mixing operator is non-renormalizable. 
Bound states in this model are supported by the exchange of one of the SU(3)$_D$ gluons, which acquires a relatively small mass during the symmetry breaking. 

The dark sector contains a triplet of Dirac fermions $\chi = (\chi_1, \chi_2, \chi_3)$ charged under SU(3)$_D$, with a Dirac mass,~$m_\chi$. After symmetry breaking, the components acquire a small mass splitting, so that $m_{\chi_1} < m_{\chi_2} = m_{\chi_3}$, with $\chi_1$ and $\chi_2$ ultimately forming an $s$-wave, spin-1 bound state, $\mathcal{B}$, which can be produced at colliders. $\chi_1$, being the lightest fermion in this theory, serves as our DM candidate.

The SU(3)$_D$ breaking occurs via three Higgs-like fields: two scalars in the adjoint representation of SU(3)$_D$, $H_1$ and $H_2$, and another scalar in the fundamental, $H_8$. The dark sector Lagrangian is given by
\begin{multline}
    \mathcal{L}_{\text{dark}} = \sum_{1,2} \frac{1}{2} D_\mu H_i^a D^\mu H_i^a + \frac{1}{2}|D_\mu H_8|^2 - V(H_1, H_2, H_8) \\
    + \overline{\chi} \left(i \slashed{D} - m_\chi \right)\chi - \frac{1}{4} V^{\mu\nu} V_{\mu\nu},
\end{multline}
where $V_{\mu\nu}$ is the SU(3)$_D$ field strength of the dark gluons, $a = 1, \cdots , 8$ is an SU(3)$_D$ index and $\tau^a \equiv \lambda^a/2$ with $\lambda^a$ being the Gell-Mann matrices. $D_\mu \equiv \partial_\mu - i g_D V_\mu^a \tau^a$ for fields in the fundamental and $D_\mu H_i^a \equiv \partial_\mu H_i^a + g_D f^{abc} V_\mu^b H_i^c$ for the two adjoint Higgs fields. 
The structure of the Gell-Mann matrices is such that $V^1$, $V^2$ couple $\chi_1$ to $\chi_2$, $V^4$, $V^5$ couple $\chi_1$ to $\chi_3$, and $V^6$, $V^7$ couple $\chi_2$ to $\chi_3$; $V^3$ couples diagonally to $\chi_1$ and $\chi_2$, while $V^8$ couples diagonally to all three fermions; the interaction vertices are shown in Appendix~\ref{app:triplehiggs}. The scalar potential  $V(H_1, H_2, H_8)$ can be chosen to satisfy the symmetry breaking pattern that we will describe below.  

We impose a $\mathbb{Z}_2$ symmetry at the renormalizable level under which $H_{1,2}^a \to - H_{1,2}^a$. This forbids any marginal interaction terms between the Higgs sector and the fermion sector, including a Yukawa interaction term. Therefore, we can treat both sectors as decoupled to first order. However, the following dimension-5 operator is allowed:
\begin{alignat}{1}
    \mathcal{L}_{\text{mass}} = \frac{1}{\Lambda_m} \left(H_8^\dagger \tau^a H_8 \right) \left(\overline{\chi} \tau^a \chi \right),
\end{alignat}
so that after $H_8$ acquires a suitable VEV, a mass splitting occurs among the components of $\chi$. Finally, we introduce the following operators that encapsulate the mixing of the dark sector with the SM:
\begin{alignat}{1}
    \mathcal{L}_{\text{mix}} = - \frac{1}{\Lambda_1} H_1^a V_{\mu\nu}^a B^{\mu\nu} - \frac{1}{\Lambda_8^2} \left(H_8^\dagger \tau^a H_8 \right) V^a_{\mu\nu} B^{\mu\nu} 
\end{alignat}
Notice that the first term introduces a small breaking of the $\mathbb{Z}_2$ symmetry. This term can originate from a dimension-6 operator that respects this discrete symmetry, such as $\phi H_1^a V_{\mu\nu}^a B^{\mu\nu}$, with $\phi$ being a scalar field that is odd under $\mathbb{Z}_2$, which acquires a VEV as well. 
The details of the origin of this operator are unimportant, as we will focus instead on the phenomenology resulting from the kinetic mixing.\footnote{One can in principle include the interaction term $H_2^a V_{\mu\nu}^a B^{\mu\nu}$, but this term does not affect the main features of this model. With the symmetry breaking pattern discussed later, the gauge bosons $V^1$ and $V^2$ couple to the same dark fermions, $\chi_1$ and $\chi_2$. We will leave this term out from the Lagrangian for simplicity.} 

At the point of symmetry breaking, $H_1$ and $H_2$ acquire a VEV $v_1$ and $v_2$ in the 1- and 2-component respectively, and $H_8$ acquires a VEV given by $\langle H_8 \rangle = v_8 (\cos \theta, 0, \sin \theta)$, with $v_8 \lesssim m_\chi \ll v_1, v_2$ and some arbitrary angle $\theta$. This symmetry breaking pattern can be accomplished by choosing an appropriate Higgs potential, which we discuss in detail in Appendix~\ref{app:triplehiggs}. The VEV of $H_1$ in the first term of $\mathcal{L}_{\text{mix}}$ leads to the conventional kinetic mixing term discussed above, with $\epsilon \equiv 2 v_1/\Lambda_1$, and $V^1$ as the mediator to the SM. The second term in $\mathcal{L}_{\text{mix}}$ guarantees the prompt decay of the other dark gluons through small mixings into the SM: details are discussed further in Appendix~\ref{app:triplehiggs}. The choice of $\langle H_8 \rangle$ gives a small mass splitting to the Dirac fields in $\chi$, leading to the following fermion masses: 
\begin{align}
    m_{\chi_1} &= m_\chi - \frac{v_8^2}{3 \Lambda_m}\, ,  \nonumber \\
    m_{\chi_2} &= m_{\chi_3} = m_\chi + \frac{v_8^2}{6 \Lambda_m} \, .
\end{align}
We will always neglect the mass splitting when not considering its role in suppressing the direct detection of DM, so that $m_{\chi_1} \simeq m_{\chi_2} = m_{\chi_3} \simeq m_\chi$. The lightest fermion $\chi_1$ is the DM candidate and it is stable; the other particles in the theory decay promptly. More details are provided in Appendix~\ref{app:triplehiggs}.  

\begin{figure}
	\centering
	\includegraphics[scale=0.3]{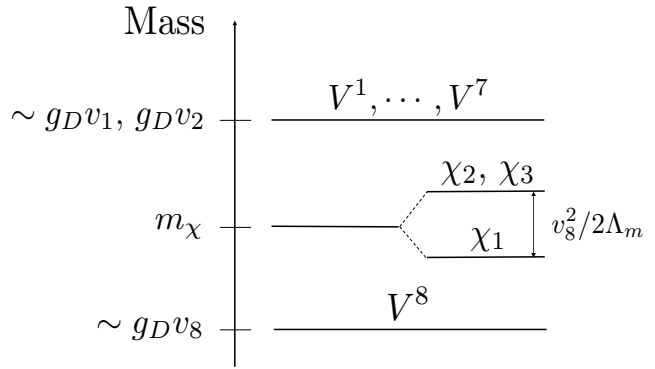}
	\caption{Spectrum of particles in the triple Higgs model.}
	\label{fig:spectrum}
\end{figure}

Finally, the dark gluons remain approximately diagonal after the symmetry breaking, with squared masses (up to order $g_D^2 v_8^2 \ll g_D^2 v_{1,2}^2$) given by:
\begin{align}
    m_1^2 &= g_D^2 v_2^2 \, , \nonumber \\
    m_2^2 &= g_D^2 v_1^2 \, , \nonumber \\
    m_3^2 &= g_D^2(v_1^2 + v_2^2) \, , \nonumber \\
    m_4^2 = m_5^2 = m_6^2 = m_7^2 &= \frac{1}{4} g_D^2 (v_1^2 + v_2^2) \, , \nonumber \\
    m_8^2 &= \frac{1}{24} g_D^2 v_8^2 (5 - 3 \cos 2 \theta) \, .
\end{align}
$m_1$ also receives $\mathcal{O}(\epsilon^2)$ corrections from the kinetic mixing with $Z$, which we will neglect as was explained above. Thus, the dark gluon masses satisfy the hierarchy
\begin{alignat}{1}
    m_8 < m_\chi < m_{1, \cdots, 7},
\end{alignat}
and $V^8$ serves as a good candidate for a bound state mediator. Fig. \ref{fig:spectrum} illustrates the spectrum of particles in this model. 

As in the Majorana case, if the dark bound state $\mathcal{B}$ arises from SM processes, then it must be produced from the mediator $V^1$; the resulting bound state must be $\overline{\chi}_1 \chi_2$ or its antiparticle equivalent $\overline{\mathcal{B}}$. Again, since the mediator is spin-1, $s$-wave bound states must be in the spin-triplet configuration. 

In the mass basis, the interaction term responsible for the production is (all fields now denote their mass eigenstate)
\begin{alignat}{1}
    \mathcal{L} \supset \frac{g_D}{2} \overline{\chi}_1 \gamma^\mu \left[V_\mu^1 - r^2 \frac{\epsilon s_W}{1 - r^2 } Z_\mu \right] \chi_2 + \text{h.c.}
\end{alignat}
with $r \equiv m_Z/m_1$. With $m_8 < m_\chi$ and the other gluons being significantly more massive than $m_\chi$, $\mathcal{B}$'s are mediated by $V^8$ through the interaction terms
\begin{alignat}{1}
    \mathcal{L}_{\mathcal{B}} = \frac{g_D}{2 \sqrt{3}} \gamma^\mu \left(\overline{\chi}_1 V^8_\mu \chi_1 + \overline{\chi}_2 V^8_\mu \chi_2 \right),
    \label{eqn:V8chichi}
\end{alignat}
which leads to an attractive potential between the constituents of $\mathcal{B}$. The coupling between $V^8$ and the fermions in $\mathcal{B}$ is therefore $\alpha_{\mathcal{B}} = \alpha_D/12$. The mass hierarchy of this model forbids decays into any of the dark gluons $V^a$ for $a = 1,\cdots,7$. Furthermore, the decay of $\mathcal{B}$ into any number of $V^8$ is forbidden by the conservation of the SU(3)$_D$ color charge in the unbroken SU(3): $V^8$ only couples $\chi_1$ to $\chi_1$ and likewise $\chi_2$ to $\chi_2$, and cannot carry away the net color charge of $\mathcal{B}$. 

\section{Experimental Constraints on Dark Sector Models}
\label{sec:experimentalConstraints}

In this section, we will first discuss in Sec.~\ref{sec:modelParameters} the range of viable model parameters in each of the dark sector models detailed above. We will then study the phenomenology of each of these models at the LHC in Sec.~\ref{sec:LHC}, and their cosmology and indirect detection signatures in Sec.~\ref{sec:id}.

\subsection{Viable Model Parameters}
\label{sec:modelParameters}

In both models, the bound state $\overline{\chi}_1 \chi_2$ (or its antiparticle equivalent, if applicable) is formed from a stable dark matter candidate $\chi_1$ and an unstable fermion $\chi_2$. In order for the decay of $\chi_2$ to not dilute the production of the bound state, we must ensure that the decay width of $\chi_2$ is much smaller than the decay width of $\mathcal{B}$. In both models, $\chi_2$ decays through an off-shell SM mediator to $\chi_1$ and two SM particles. In the pseudo-Dirac model, this three-body decay width is parametrically $\Gamma_{\chi_2} \sim \epsilon^2 g_D^2 g_{\text{SM}}^2 (\Delta m)^5/m_V^4$, where $g_{\text{SM}}$ is a coupling constant to the SM which depends on the actual SM particle considered, and $\Delta m=m_{\chi_2}-m_{\chi_1}$, which we always take to be small. On the other hand, the bound state decay width is $\Gamma_{\mathcal{B}} \sim \epsilon^2 g_D^2 g_{\text{SM}}^2 m_\chi^2 |\psi(0)|^2 / m_V^4$. The relative ratio of these widths is therefore
\begin{alignat}{1}
    \frac{\Gamma_{\chi_2}}{\Gamma_{\mathcal{B}}} \sim \frac{(\Delta m)^5}{\alpha_{\mathcal{B}}^3 m_\chi^5} \ll 1,
    \label{eqn:Gammachi2OverGammaB}
\end{alignat}
where $\alpha_\mathcal{B} = y_D^2/4 \pi \sim \mathcal{O}(0.1 - 1)$ for situations where LHC production of bound states is important. An identical relationship holds for the triple Higgs model, with $\alpha_{\mathcal{B}} = \alpha_D/12$. 

In both of the models we have presented, the interaction between the dark sector and the SM is controlled by a single vector boson: $V$ in the pseudo-Dirac model of Sec.~\ref{sec:PseudoDirac} and $V^1$ in the triple Higgs model of Sec.~\ref{sec:TripleHiggs}. 
The mixing of the SM and dark sectors shifts the $Z$ mass, and is thus constrained by EW precision tests~(EWPT). In particular, the $\rho$ parameter is shifted by an amount~\cite{Cassel:2009pu}
\begin{alignat}{1}
    \Delta \rho = -\frac{m_W^2}{m_V^2} t^2_W \cdot \epsilon^2 + \mathcal{O}\left( \frac{m_W^4}{m_V^4} \right),
\end{alignat}
where $m_V$ is the mass of the SM mediator in either model, and $t_W$ is the tangent of the weak mixing angle. The global fit for the central value of $\rho$ to EWPT data is $\rho_0 = 1.00037 \pm 0.00023$~\cite{Patrignani:2016xqp}. Constraints are set by requiring that any choice of $\epsilon$ leads to a minimum value of $m_V$ such that $\Delta \rho$ is consistent with the 2$\sigma$ limit for the value of $\rho_0$. 

Next, in order for a bound state to be possible, the constraint given in Eq.~\eqref{eqn:boundStateRequirement} must be satisfied. This condition can be satisfied by ensuring that the mass of the particle supporting the bound state is sufficiently small. For the pseudo-Dirac model, this means choosing a sufficiently small dark Higgs mass such that $y_D^2 m_\chi > 21.1 m_{h_D}$, and for the triple Higgs model, ensuring that $\alpha_D m_\chi > 20.16 m_8$.

Finally, to avoid direct detection constraints, the mass splitting must exceed the typical kinetic energy of DM in the solar circle. Taking the velocity dispersion of DM to be $v \sim 10^{-3}$, this means that the mass splitting has to exceed approximately $10^{-6} m_\chi$. A small mass splitting, albeit large enough to be consistent with this lower bound, can be achieved by picking suitable values for the Dirac bare mass $m_D$ in the pseudo-Dirac model and $\Lambda_m$ in the triple Higgs model. 

In both theories, there are two parameters ($m_D$ and $m_{h_D}$ for the pseudo-Dirac model, $m_8$ and $\Lambda_m$ for the triple Higgs model) that can be set to naturally satisfy both the criterion for bound states and avoid direct detection constraints, while having little impact on the LHC phenomenology. However, these parameters can have some influence on the relic abundance of DM in these theories, as well as on indirect detection bounds. This will be discussed after the next section.

\subsection{LHC Phenomenology}
\label{sec:LHC}

We now turn our attention to the production and detection of bound states at the LHC for both theories. In the perturbative picture, bound states $\mathcal{B}$ are produced by quark anti-quark parton interactions through an $s$-channel $V$ and $Z$ (mass-eigenstate) boson, with the only available decay mode of $\cB$ being an off-shell $V$ or $Z$ back into SM particles, leading to resonance signatures. The more accurate procedure of taking into account the mixing of $V$ and $\mathcal{B}$ yields a qualitatively similar result; we use the full mixing calculation in all of the plots shown, but focus our qualitative discussion primarily on the perturbative picture.\footnote{We neglect any mixing between $V$, $\mathcal{B}$ with $Z$, since we will usually take $V$ and $B$ to be much heavier than $Z$, and the coupling between $Z$ and the dark sector particles is suppressed by $\epsilon$.}

The mono-$X$ + MET search can be effective in setting constraints on these dark models, particularly in the range of parameter space where $2m_\chi < m_V$, the region of interest for both dark sector models. To study the constraints that mono-jet + MET searches can place on our models, we use \texttt{FeynRules}~\cite{Alloul:2013bka} and \texttt{MadGraph}~\cite{Alwall:2014hca} to obtain the MET distribution for a wide range of $m_\chi$ and $m_V$. The distribution is then compared to the observed 95\,\% confidence upper limit on the number of mono-jet + MET events in 10 inclusive MET bins obtained by ATLAS with \SI{36.1}{fb^{-1}} of data~\cite{Aaboud:2017phn}. Any value of $m_\chi$ and $m_V$ with a MET distribution that has more events in any inclusive bin than the 95\% upper limit is deemed to be ruled out by the experiment. 

Next, we recast bounds from a search for resonance in dilepton events in \SI{36.1}{fb^{-1}} of 13\,TeV ATLAS data~\cite{ATLAS:2017wce} to set constraints on the production of $\mathcal{B}$. 
In the models considered here, $\mathcal{B}$ decays entirely into SM particles with a significant branching ratio to pairs of leptons, making the dilepton resonance search a particularly powerful probe. This search constrains the production cross section times branching ratio of a $Z'$ boson assuming some minimal vector couplings to the SM fermions, which allows us to directly interpret these constraints as a limit on the production of cross section times branching ratio of the bound state $\mathcal{B}$. 

These searches are also sensitive to the resonant production of the vector mediator $V$ itself, which tends to be significantly more constraining than mono-jet + MET searches when the coupling of the mediator to SM quarks are comparable to the coupling to DM. However, in portal models like the ones we are considering, the mixing into the SM $\epsilon$ is small while the coupling to DM $\alpha_D$ can be large. In the range of parameter space where the mediator mass $m_V \gtrsim 2m_\chi$, $V$ overwhelmingly decays into $\overline{\chi}_1 \chi_2$ or $\overline{\chi}_2 \chi_1$, which correspond to final states with MET and are vetoed in dilepton resonance searches to suppress $W$ and $Z$ backgrounds~\cite{ATLAS:2017wce}. The search for $\mathcal{B}$, however, faces no such limitation in this region of parameter space.

The production cross section of $\mathcal{B}$ (and equivalently of $V$) can be computed from Eq.~\eqref{eqn:qqbartoB}, assuming the narrow width approximation. 
In the perturbative picture, $\mathcal{B}$ decays through an $\epsilon$-suppressed coupling to the $Z$, or through $V$, which has an $\epsilon$-suppressed coupling to both $J^\mu_{\text{EM}}$ and $J^\mu_Z$. The resulting expression for the bound state width to quarks is
\begin{multline}
    \label{eqn:Bwidth}
    \Gamma_{\mathcal{B} \to q \overline{q}} = \frac{16 \pi N_m r^4 m_\chi^2}{c_W^2(m_Z^2 - 4 m_\chi^2)^2} \\
    \times \frac{\alpha \alpha_D \epsilon^2 |\psi(0)|^2}{(m_Z^2 - 4 r^2 m_\chi^2)^2 + r^2 m_Z^2 \Gamma_V^2(s = m_\mathcal{B}^2)} \\
    \times \left[ \left(c_W^2 Q(m_Z^2 - 4m_\chi^2) + 4 g_V m_\chi^2\right)^2 + 16 g_A^2 m_\chi^4  \right. \\
    \left. + \frac{r^2}{(1 - r^2)^2}  \Gamma_V^2(s = m_\mathcal{B}^2)(g_V^2 + g_A^2) \right] \,,
\end{multline}
%
%
%
where $\alpha$ is the EM fine structure constant, $Q$ is the electric charge of the quark, $g_V$ and $g_A$ are the vector and axial couplings of $q$ to the $Z$ respectively, given by  $g_V = \{0.25, -0.0189, 0.0959, -0.1730 \}$ and $g_A = \{0.25, -0.25, 0.25, -0.25\}$ for $\{\nu_e, e, u, d\}$ and for the other 2 generations respectively. $N_m = 4$ for the pseudo-Dirac model and $N_m = 1$ for the triple Higgs model, which accounts for the difference in coupling and fermion types. As previously, $|\psi(0)|^2$ is the squared amplitude of the wave function of the bound state at the origin, given explicitly by
\begin{alignat}{1}
    |\psi(0)|^2 = \begin{cases}
        \left( \frac{y_D^2}{4\pi} \right)^3 \frac{m_\chi^3}{8 \pi}, &\text{Pseudo-Dirac}, \\
        \left(\frac{\alpha_D}{12}\right)^3 \frac{m_\chi^3}{8\pi},  &\text{Triple Higgs}.
    \end{cases}
    \label{eqn:psi02}
\end{alignat}
Note that we have assumed throughout that the bound state is well-approximated by non-relativistic quantum mechanical results, which is a valid assumption so long as the binding energy of $\mathcal{B}$ is far less than $m_\chi$. For this bound state, we thus require
\begin{alignat}{1}
    \frac{1}{4}\alpha_\mathcal{B}^2 m_\chi \ll 2 m_\chi,
\end{alignat}
where $\alpha_\mathcal{B} = y_D^2/4\pi$ for the pseudo-Dirac case, and $\alpha_\mathcal{B} = \alpha_D/12$ for the triple Higgs model. 

As we argued earlier, the production cross section of $\mathcal{B}$ crucially depends on the total width of $V$; this means that the total width of $V$ should be included in the computation of the width shown in Eq.~\eqref{eqn:Bwidth}. Importantly, the width of $V$ should be evaluated at $s = m_\mathcal{B}^2$, since $\mathcal{B}$ lies below the $\chi \overline{\chi}$ open production threshold \cite{Patrignani:2016xqp}. The perturbative partial widths of $\mathcal{B}$ as well as $V$ into all possible SM final states are shown in Appendix~\ref{app:widths}. 

In the mixing picture, the partial widths of $V$ calculated here correspond to $\Gamma_{V,0}$. We take $\Gamma_{\mathcal{B},0} = 0$, since $\Gamma_{\mathcal{B},0} = \Gamma_{\chi_2} \ll \Gamma_{\mathcal{B}}$, as shown in Eq.~(\ref{eqn:Gammachi2OverGammaB}). In the pseudo-Dirac model, there is only one bound state, and the mixing calculation proceeds in the same fashion as described in Sec.~\ref{sec:VBMixing}. The sum of the perturbative partial widths of $\mathcal{B}$, calculated in Appendix~\ref{app:widths}, is numerically a good approximation to the width after mixing, $\Gamma_{\mathcal{B}}$. For the triple Higgs model, there are two bound states, $\mathcal{B}$ and $\overline{\mathcal{B}}$, and so all three states need to be simultaneously diagonalized. However, $\mathcal{B}$ and $\overline{\mathcal{B}}$ maximally mix to form two CP eigenstates,
\begin{alignat}{1}
    \mathcal{B}_\pm = \frac{\ket{\mathcal{B}} \pm \ket{\overline{\mathcal{B}}}}{\sqrt{2}}.
\end{alignat}
Since $V^1$ is a CP-even state, it does not mix with the CP-odd combination $\mathcal{B}_-$, and the diagonalization is performed over $V^1$ and the CP-even $\mathcal{B}_+$; the CP-odd state $\mathcal{B}_-$ does not interact with the SM.
In both models, the unmixed mass matrix given in Eq.~(\ref{eqn:massMixing}), with the mixing parameter $f$ given by \cite{Franzini:1987jw,Kuhn:1985eu,Kuhn:1987ty}
\begin{alignat}{1}
    f = 4 N_f \psi(0) \sqrt{\pi \alpha_D m_{\mathcal{B},0}}\, ,
\end{alignat}
where $N_f = 1$ for the pseudo-Dirac model, and $N_f = 1/\sqrt{2}$ for the triple Higgs model, which accounts for the differences in coupling and bound-state mixing.

In both models, $\mathcal{B}$ cannot decay into final states that only contain the mediator which supports the bound state: this is because both the dark Higgs in the pseudo-Dirac model and $V^8$ in the triple Higgs model have couplings with the DM fermion number that conserves the number of each of $\chi_1$ and $\chi_2$. 

Decays of $\mathcal{B}$ into dark sector final states become possible once $m_{\mathcal{B}} \gtrsim m_V$ in the pseudo-Dirac model, or $m_{\mathcal{B}} \gtrsim m_1$ in the triple Higgs model: the final states are $Vh_D$ and $V^1 V^8 V^8$ respectively. Because of the large coupling between the DM fermions and the mediators, these dark sector decays are the main decay modes of $\mathcal{B}$, rendering the dilepton resonance search for $\mathcal{B}$ ineffective. These dark sector final states all mix with the SM, and can in principle lead to multilepton signatures at the LHC. Earlier studies have exploited this signature to look for bound states \cite{An:2015pva,Bi:2016gca}, but we do not explore this possibility here for two reasons. First, once $m_{\mathcal{B}}$ becomes significantly greater than the SM mediator mass, the resonant enhancement derived in Eq.~(\ref{eqn:qqbartoBlimit}) becomes ineffective, and the cross section for producing $\mathcal{B}$ drops quickly away from $m_{\mathcal{B}} \sim m_V$ or $m_1$. Furthermore, the branching ratio of these mediators to leptons is small, since they kinetically mix through the U(1)$_Y$ and decay predominantly into quarks. Second, a direct search for the $V$ or $V^1$ resonance is significantly more constraining, since the mediator is lighter than the bound state, and there is one fewer factor of the branching ratio to leptons to contend with. In both models, the mediator dilepton resonance search rules out all of the parameter space for $m_\mathcal{B} > m_1$ or $m_V$ once the coupling to the SM is sufficiently large.

At tree level, we are therefore only interested in the decay modes of $\mathcal{B}$ and $V$ into the SM: both particles can decay into a pair of SM fermions, as well as $W^+W^-$ and $Zh$ where $h$ is the SM Higgs, through the mixing of $V$ with $Z/\gamma$. Neither particle can decay into $ZZ$ or $\gamma \gamma$ final states, since these processes are forbidden by charge conjugation symmetry. 

The sensitivity of the dilepton resonance search depends strongly on the width of the resonance, and the 13 TeV ATLAS limits with \SI{36.1}{fb^{-1}} of data as a function of the ratio of the width of the resonance to its mass $\Gamma/m$ are presented in \cite{ATLAS:2017wce}. The total widths of both states are fully taken into account when computing the limits of the search, and the search is assumed to be completely ineffective once $\Gamma/m > 0.32$.

\begin{figure*}[t!]
	\subfloat[]{
		\label{fig:LHCPseudoDirac}
		\includegraphics[scale=0.58]{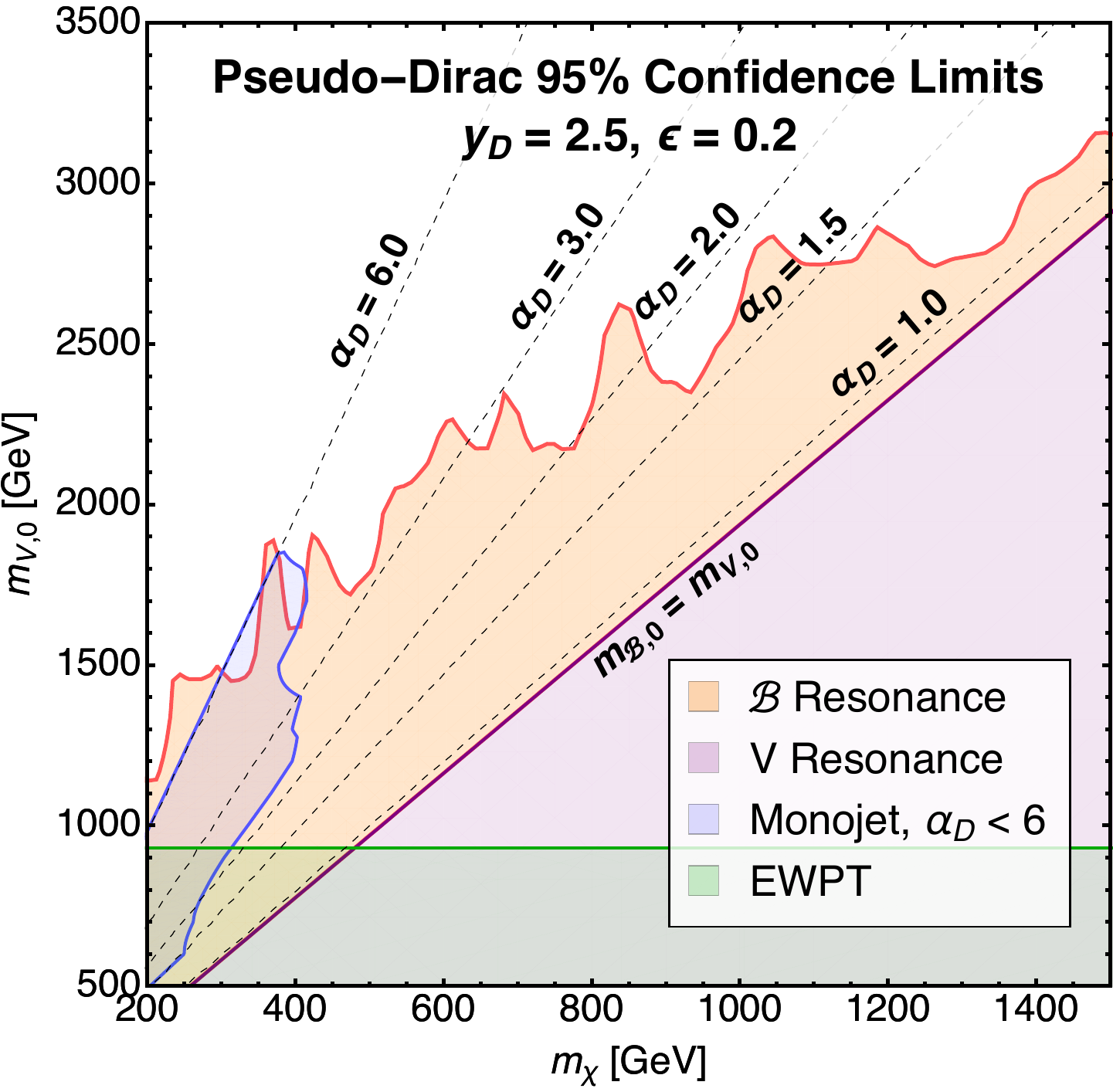}
	}
	\hfil
	\subfloat[]{
		\label{fig:LHCTripleHiggs}
		\includegraphics[scale=0.58]{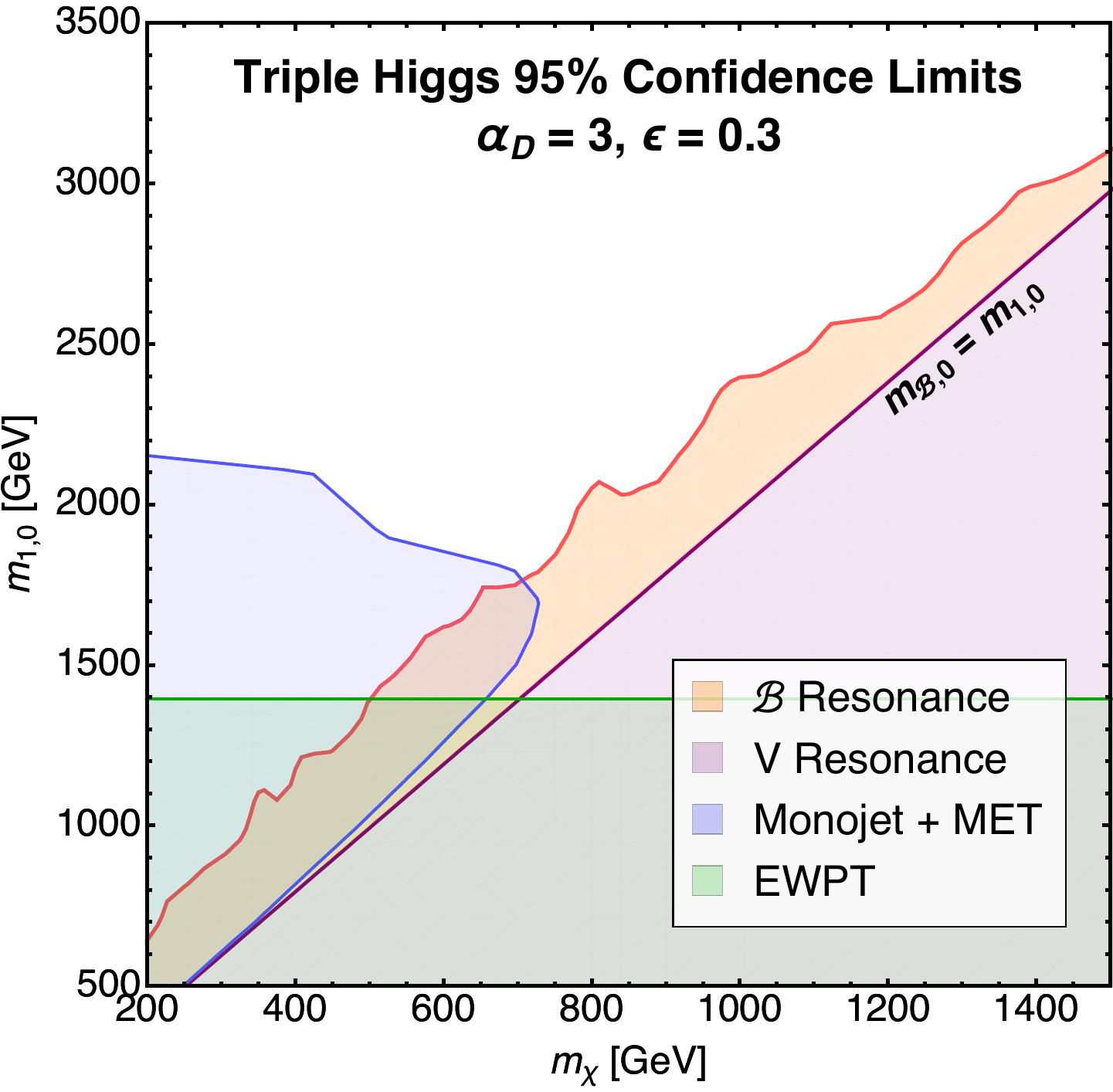}
	}
  \caption{95\% confidence limits in the $m_\chi - m_{V,0}$ plane of the pseudo-Dirac model~(left) and in the $m_\chi - m_{1,0}$ plane for the triple Higgs model~(right). $m_{V,0}$ and $m_{1,0}$ are the unmixed masses of the mediator in each respective model. All resonance calculations are made using the full mixing calculation.
  Experimental constraints from mono-jet + MET~(blue), dilepton $\mathcal{B}$ resonance~(orange), dilepton $V$ resonance~(purple) and EWPT constraints~(green) are shown for $y_D = 2.5$, $\epsilon = 0.2$ for the pseudo-Dirac model, and $\alpha_D = 3$, $\epsilon = 0.3$ for the triple Higgs model. All dilepton resonance searches are for the lighter mass eigenstate after mixing. For the pseudo-Dirac model on the left, the dark sector coupling $\alpha_D$ is completely fixed by a choice of $\{m_\chi, m_V, y_D\}$; contours (black, dashed) indicate the value of $\alpha_D$ on the $m_\chi - m_V$ plane when $y_D = 2.5$.
  }
  \label{fig:LHC}
\end{figure*}

\begin{figure*}[t!]
	\subfloat[]{
		\label{fig:LHCPseudoDirac2}
		\includegraphics[scale=0.58]{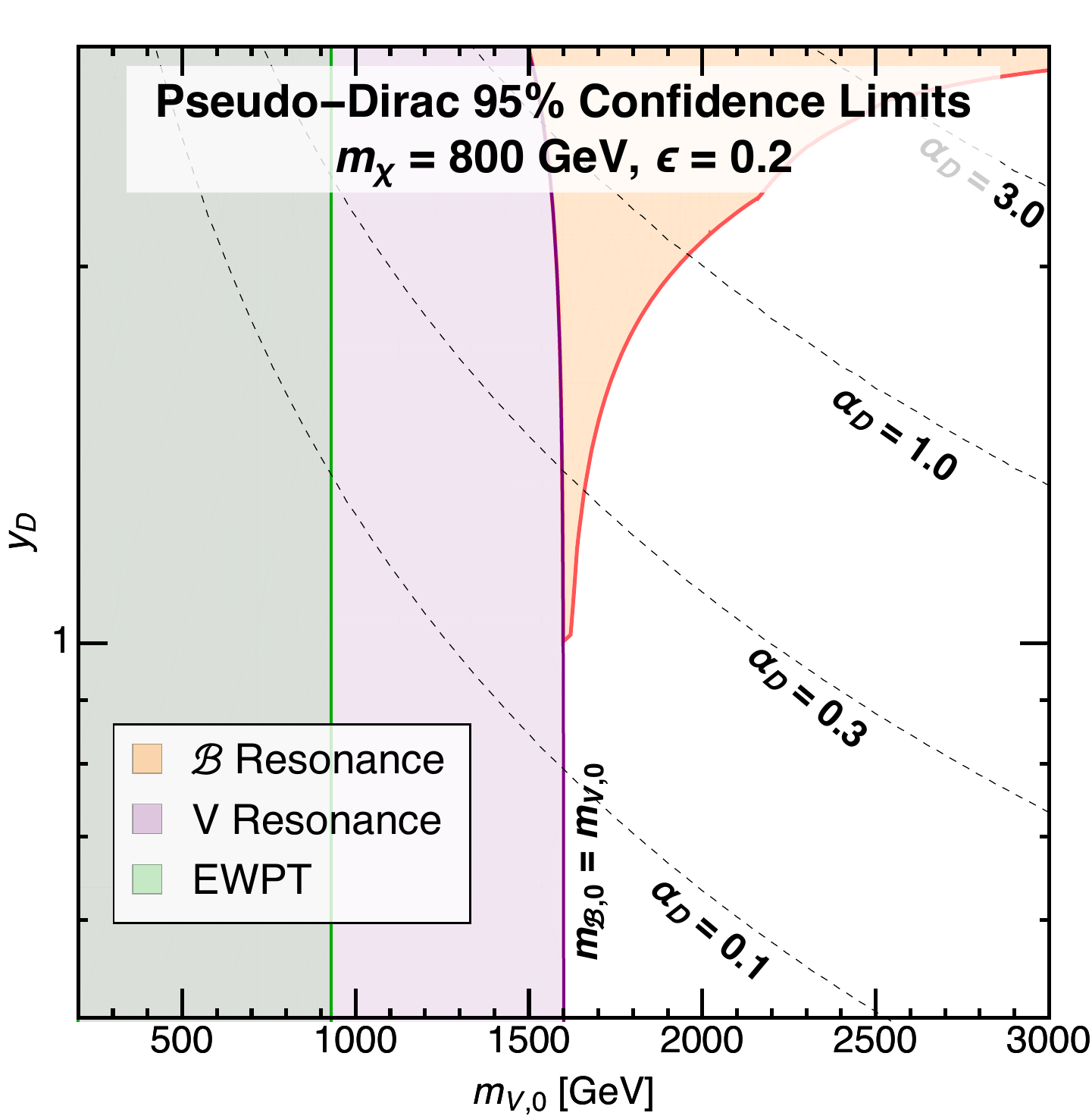}
	}
	\hfil
	\subfloat[]{
		\label{fig:LHCTripleHiggs2}
		\includegraphics[scale=0.59]{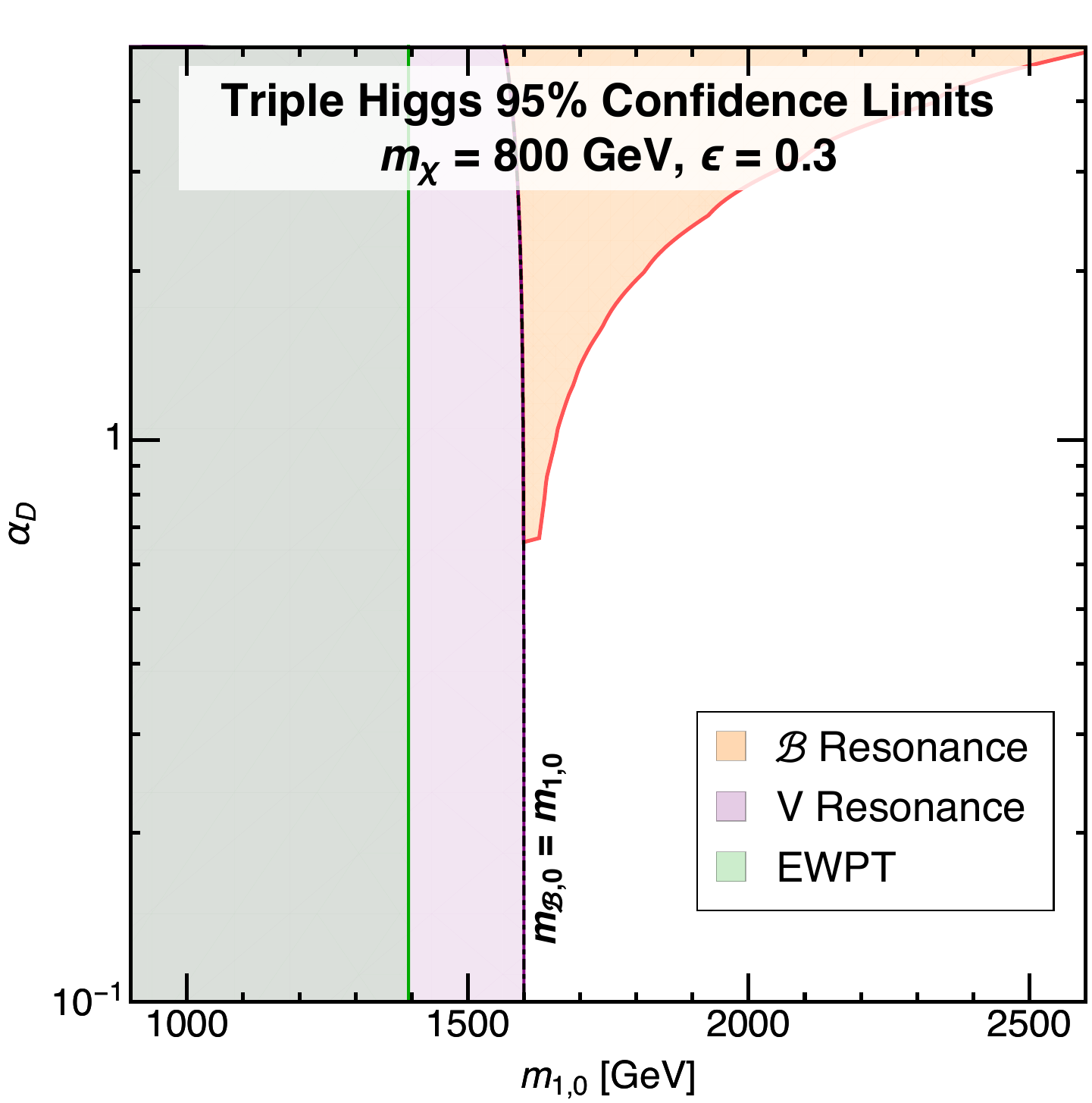}
	}
  \caption{95\% confidence limits in the $m_{V,0} - y_D$ plane of the pseudo-Dirac model~(left) and in the $m_{1,0} - \alpha_D$ plane of the triple Higgs model~(right), similar to Fig.~\ref{fig:LHC}. Experimental constraints from dilepton $\mathcal{B}$ resonance~(orange), dilepton $V$ resonance~(purple) and EWPT constraints~(green) are shown for $m_\chi = 800$ GeV, $\epsilon = 0.2$ for the pseudo-Dirac model, and $m_\chi = 800$ GeV, $\epsilon = 0.3$ for the triple Higgs model. Contours (black, dashed) of $\alpha_D$, which is fixed for a given choice of $\{m_\chi, m_V, y_D\}$ for the pseudo-Dirac model, are also shown. 
  }
  \label{fig:LHC2}
\end{figure*}

The resulting 95\% confidence limits from mono-jet + MET, dilepton resonance and EWPT are shown in Fig.~\ref{fig:LHC} in the $m_\chi - m_V$ plane for the pseudo-Dirac model and in Fig.~\ref{fig:LHC2} in the $m_\chi - \alpha_D$ plane for the triple Higgs model.

The dilepton resonance search results presented in both figures are searches for the lighter resonance state in the mixing picture; the search switches from $V$ to $\mathcal{B}$ along the line $m_{V,0} = m_{\mathcal{B},0}$, where the lighter resonance changes rapidly from being mostly $V_0$ to mostly $\mathcal{B}_0$ as one moves from below to above this line.\footnote{In spite of this, the partonic cross section including both $V$ and $\mathcal{B}$ is continuous across this line; it is only the particle that should be identified with the narrow Breit-Wigner signal at the low mass eigenvalue that changes.} As we argued earlier, since the mass eigenstates are always well-separated and the lighter resonance is always narrow, we can simply assume that the total cross section is given by a Breit-Wigner profile with a width given by either the $V$ ($m_{V,0} < m_{\mathcal{B},0}$) or the $\mathcal{B}$ ($m_{\mathcal{B},0} > m_{V,0}$) and neglect interference effects. In both cases, the search for the $\mathcal{B}$ resonance when $m_{V,0} > m_{\mathcal{B},0}$ extends the reach of experimental constraints significantly into this region of parameter space, as compared to what we might expect from just the vector resonance search and the mono-jet + MET search combined.


\subsection{Freezeout and Indirect Detection}
\label{sec:id}

We now turn our attention to the freezeout process for the DM in each model, as well as constraints derived from indirect detection experiments. Let us focus on the annihilation channels that do not suffer a suppression by $\epsilon$, in order to be as model-independent as possible. In the pseudo-Dirac model, the potential kinematically available final states (at late times) are $h_D h_D$ and $\mathcal{B}^\prime h_D$, with the latter channel corresponding to radiative formation of a bound state, $\mathcal{B}^\prime$ (which may be spin-1 or spin-0). The $Vh_D$ final state is forbidden, since $V$ couples $\chi_1$ to $\chi_2$, and $h_D$ couples $\chi_1$ to $\chi_1$. In the triple Higgs model, if all the gauge bosons and Higgses except $V^8$ are heavier than the DM, the only open final states are $V^8 V^8$ and the radiative bound state formation. Note that in the limit where the DM is slow-moving, radiative bound state formation requires not merely that the mediator be light compared to $\alpha_\cB m_\chi$, as required for a bound state, but that it satisfy the stronger condition that the mediator mass is smaller than the binding energy, $m_Y \lesssim \alpha_\cB^2 m_\chi/4$. 
Thus, this process can be forbidden by increasing the mediator mass, and indeed we will see that indirect detection limits are much easier to satisfy in regions of parameter space where $\alpha_\mathcal{B} m_\chi \gtrsim m_Y \gtrsim \alpha_\mathcal{B}^2 m_\chi/4$. In this regime, the DM annihilation products will thus be determined by the decays of the bound state mediator.

During freezeout, the partner particles $\chi_2$ (in the pseudo-Dirac model) and $\chi_2$, $\chi_3$ (in the triple Higgs model) are also present, and their annihilation and co-annihilation channels may also relevant.

\begin{figure*}[t!]
	\subfloat[]{
		\label{fig:indirectPseudoDirac}
		\includegraphics[scale=0.58]{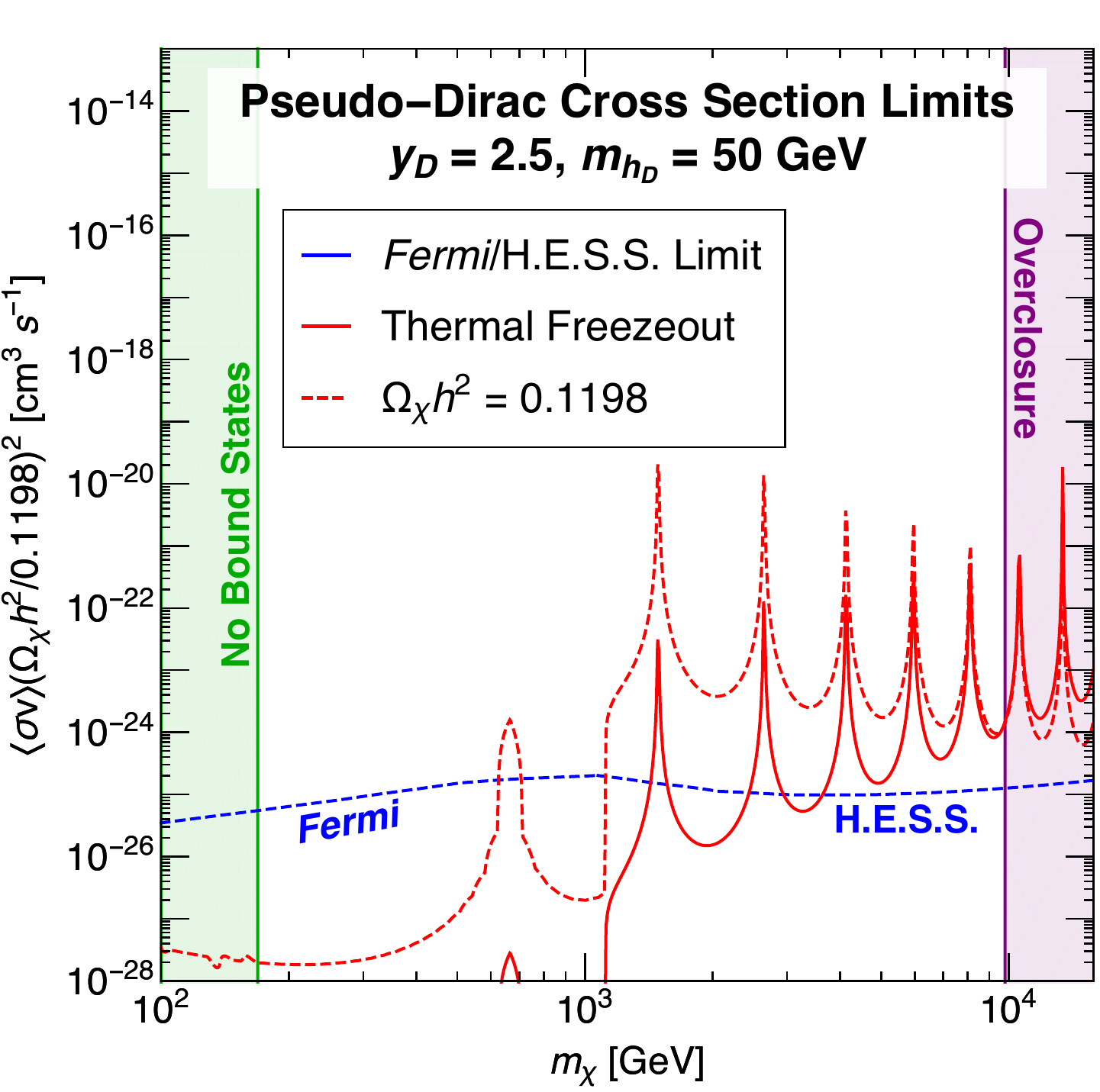}
	}
	\hfil
	\subfloat[]{
		\label{fig:indirectTripleHiggs}
		\includegraphics[scale=0.58]{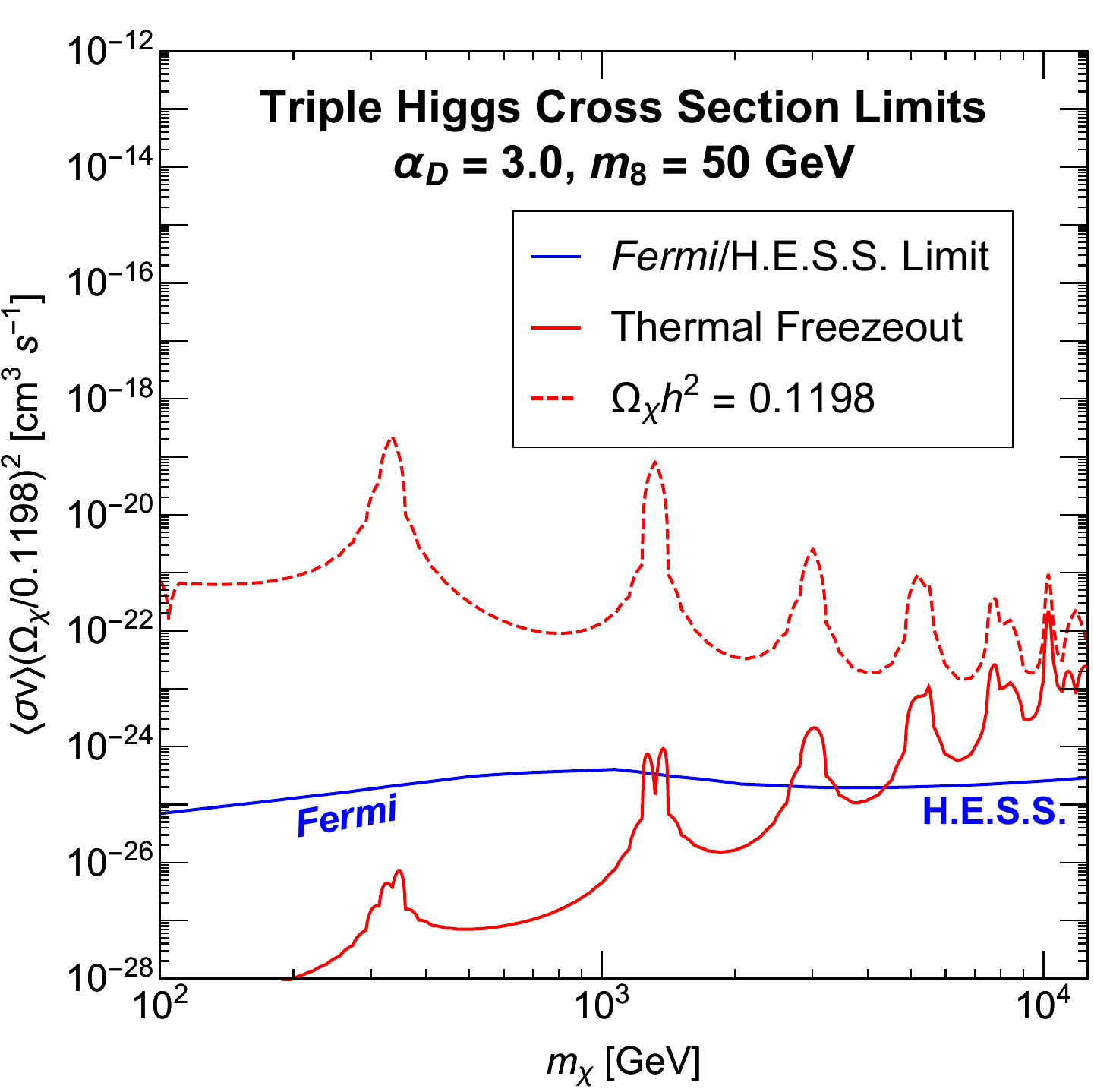}
	}
  \caption{Comparison of predicted DM annihilation rates (including Sommerfeld enhancement and radiative bound state formation) to constraints on the $\bar{b} b$ channel from \textit{Fermi} observations of dwarf galaxies~\cite{Fermi-LAT:2016uux} and H.E.S.S. observations of the Galactic center region~\cite{Abdallah:2016ygi}. The red solid line indicates the predicted cross section, rescaled by the fraction of DM squared, for thermally produced DM. For the total DM abundance we take $\Omega_\chi h^2 = 0.1198$ \cite{PlanckCollaboration2015}. The red dashed line shows the predicted cross section only, corresponding to the assumption that the annihilating species constitutes 100\% of the DM. The region to the right of the vertical purple line is ruled out by overproducing the DM abundance. The left panel shows the result for the pseudo-Dirac model with $y_D=2.5$; the right panel shows the result for the triple-Higgs model with $\alpha_D = 3.0$.}
  \label{fig:indirect1}
\end{figure*}

\begin{figure*}[t!]
	\subfloat[]{
		\label{fig:indirectPseudoDirac2}
		\includegraphics[scale=0.58]{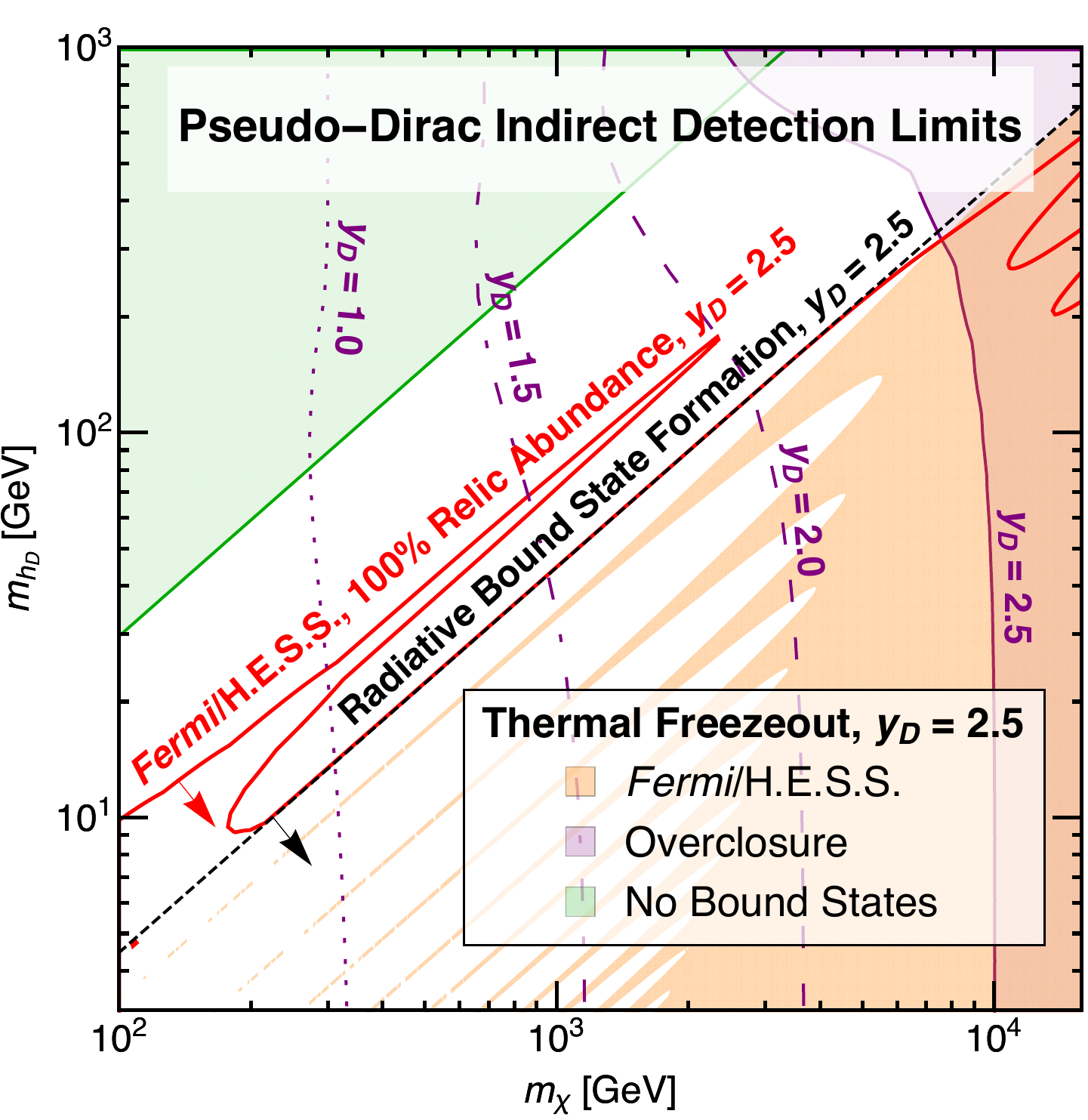}
	}
	\hfil
	\subfloat[]{
		\label{fig:indirectTripleHiggs2}
		\includegraphics[scale=0.58]{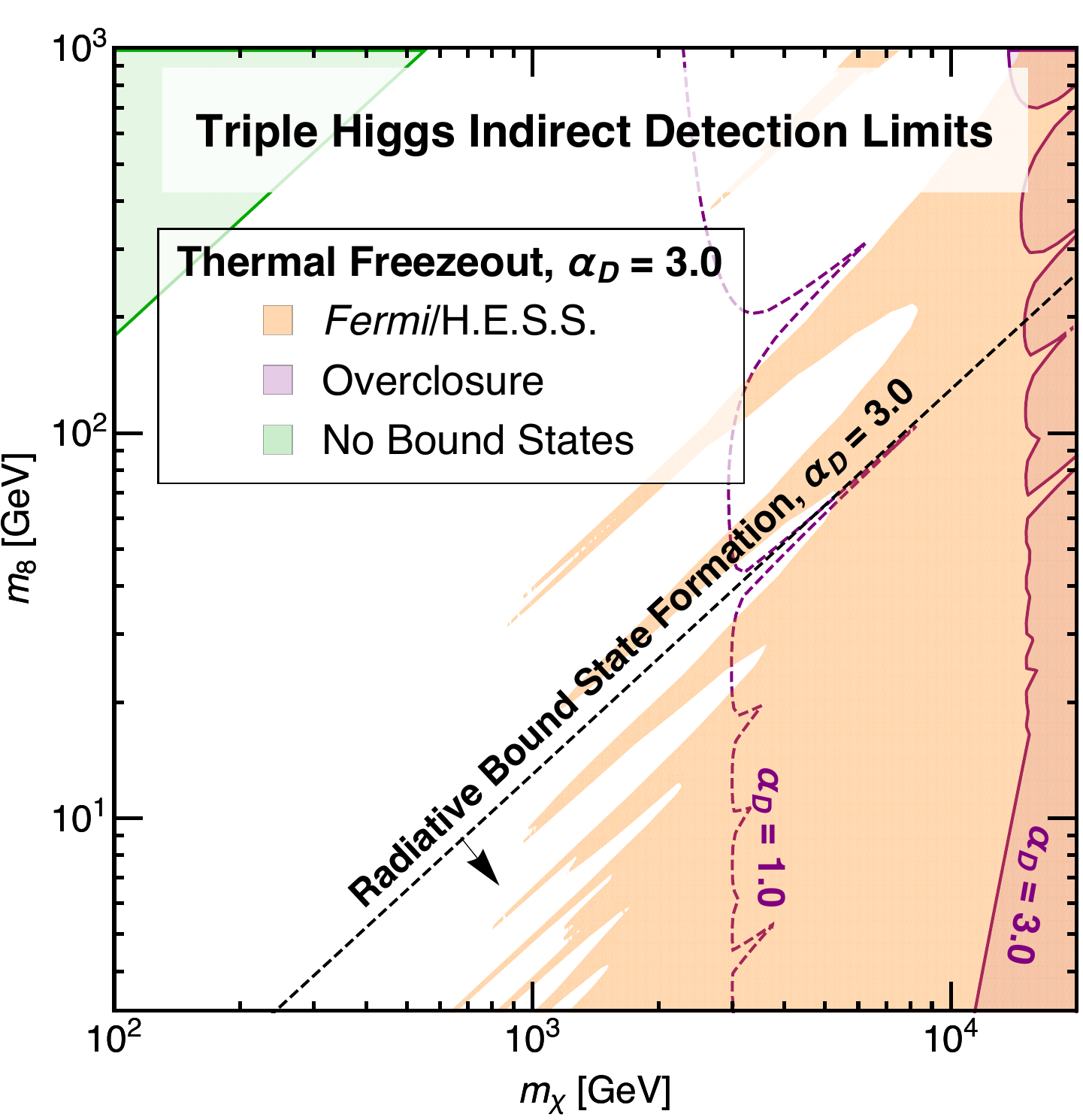}
	}
  \caption{Indirect detection and overclosure limits on the $m_\chi - m_{h_D}$ plane of the pseudo-Dirac model~(left) and the $m_\chi - m_8$ plane of the triple Higgs model~(right). Shaded regions indicate excluded regions where no bound states exist~(green), the cosmological DM abundance is overproduced~(purple), and the estimated gamma-ray signal exceeds bounds from the \textit{Fermi} and H.E.S.S. telescopes~(orange), for $y_D = 2.5$ for the pseudo-Dirac model and $\alpha_D = 3.0$ for the triple-Higgs model. In the region below the dashed black line, bound state formation can proceed in the Milky Way halo through emission of an on-shell $h_D$ or $V^8$, and contributes to the indirect detection signal. Dashed, dot-dashed and dotted purple lines indicate the more stringent overclosure limits for smaller values of the coupling. In the left panel, the region below the solid red line is excluded by gamma-ray bounds if the DM candidate of the model is assumed to be symmetric and to comprise 100\% of the DM (from a non-thermal origin). In the right panel, for the case where we assume that the DM candidate comprises 100\% of the DM, the entire parameter space for $\alpha_D = 3.0$ is excluded.
  }
  \label{fig:indirect2}
\end{figure*}

\subsubsection{Pseudo-Dirac Model}

If $m_V > 2 m_\chi$ and the bound-state mediator is too heavy for radiative bound state formation, then the only annihilation channel not suppressed by $\epsilon$ or kinematically forbidden is annihilation to $h_Dh_D$\,. Both $\chi_1 \chi_1$ and $\chi_2 \chi_2$ pairs can annihilate in this fashion, but there is no tree-level coannihilation; $\chi_1 \chi_2 \rightarrow h_D h_D$ does not occur for the same reason that the $\chi_1 \chi_2$ bound state does not decay into the dark sector. The cross section for $\chi_i \chi_i \rightarrow h_D h_D$ in the limit of low DM velocity, before accounting for the Sommerfeld enhancement, is given by:
\begin{align} \sigma v_\text{rel}
& = \frac{\pi}{6} v_\text{rel}^2 \left( \frac{ y_D^2}{ 4 \pi} \right)^2 \frac{(9 - 8x_h^2 + 2x_h^4) \sqrt{1 - x_h^2}}{(2 - x_h^2)^4 m_\chi^2},
 \end{align}

where $x_h \equiv m_{h_D}/m_\chi$. We will assume that during freezeout the mass splitting between $\chi_2$ and $\chi_1$, set by $m_D$, is small compared to the freezeout temperature; for $\mathcal{O}$(TeV) DM this corresponds to requiring a mass splitting at the GeV scale or below, which is not in tension with the requirement that the mass splitting be large enough to prevent elastic scattering in the present-day halo (where typical kinetic energies for a TeV DM particle are of order 1\,MeV or less). 
In this case, the abundances of $\chi_1$ and $\chi_2$ remain equal during freezeout, as their equilibrium abundances are equal and their annihilation channels are identical. 
Consequently, each of $\chi_1$ and $\chi_2$ must constitute half the DM abundance, with the $\chi_2$ subsequently decaying to $\chi_1$ ({this occurs through emission of an off-shell $V$).

Since $p$-wave processes can dominate during freezeout, to compute the full rate we will need the Sommerfeld enhancement factor for higher-$l$ processes. The Sommerfeld enhancement for multipole $l$ due to a Yukawa potential can be numerically approximated by~\cite{Cassel:2009wt}:
 \begin{multline}
     S_l 
     \approx 
     \frac{\pi}{\epsilon_v} \frac{\sinh\left(2\pi \delta \right)}{\cosh\left(2\pi \delta \right) - \cosh\left(2\pi \delta \sqrt{1 - \epsilon_\phi^*/\epsilon_v^2}\right)}  \\
   \times \prod_{k=1}^l  \frac{k^4 \epsilon_\phi^{*2} + 2 k^2 (2 \epsilon_v^2  - \epsilon_\phi^*) + 1}{k^2 \epsilon_\phi^{*2} + 4 \epsilon_v^2},  
 \end{multline}
 where $\delta \equiv \epsilon_v / \epsilon_\phi^*$, $\epsilon_v = v_\text{rel}/(2\alpha_\mathcal{B})$ and $\epsilon_\phi^*= (\pi^2/6) m_Y/(\alpha_\mathrm{B} m_\chi)$.
 
 We determine the relic density by numerically solving the Boltzmann equation (following the method of~\cite{Steigman:2012nb}) for the $\chi_1$ state and then doubling the result to account for the contribution from $\chi_2$. We integrate the Sommerfeld-enhanced velocity-dependent cross section over the thermal velocity distribution (assuming a Maxwell-Boltzmann distribution) for the DM at each timestep. As discussed earlier, we neglect radiative bound state formation during freezeout. We define overclosure to occur when $\Omega_\chi h^2 > 0.1228$, corresponding to the $2 \sigma$ upper limit ($0.1198 + 2 \times 0.0015$) from Ref.~\cite{PlanckCollaboration2015}.
 
To estimate the signal in indirect detection, we first compute the Sommerfeld-enhanced cross section for $\chi_1 \chi_1 \rightarrow h_D h_D$ in the Milky Way halo, assuming the local DM velocity distribution in the Galactic frame is Maxwellian, $f(v) = \sqrt{2/\pi} v^2 e^{-v^2/2\sigma^2}/\sigma^3$, with $\sigma = 150$\,km/s. This choice corresponds to $v_c = \sqrt{2} \sigma \sim 220$ km/s, consistent with the standard halo model~\cite{1986MNRAS.221.1023K,2009ApJ...700..137R,2010MNRAS.402..934M,Herzog-Arbeitman:2017zbm}.

For $m_{h_D}$ smaller than the binding energy, we also account for radiative formation of $\chi_1 \chi_1$ bound states (followed by decay into SM particles). To estimate the bound state formation rate via light scalar emission at low velocities, we add to this rate the analytic low-velocity estimate of~\cite{An:2016kie} for the cross section for capture into the ground state (which dominates the overall capture rate),
\begin{equation} 
	\sigma v_\text{rel} 
	\approx 
	\frac{1}{2} \frac{ \pi \alpha_\mathcal{B}^2}{m_\chi^2} \frac{2^6\pi^2 \alpha_\mathcal{B}^2 e^{-4}}{9 \epsilon_\phi^* \sin^2(\pi/\sqrt{\epsilon_\phi^*})} \, .
	\label{eq:scalarbound}
\end{equation}
Note that Ref.~\cite{An:2016kie} derives this expression from the Hulth\'{e}n potential, so in Eq.~\eqref{eq:scalarbound} we have replaced $m_\phi/(\alpha_D m_D)$ in their result with the parameter $\epsilon_\phi^*$; the Hulth\'{e}n potential with this rescaled mass parameter gives a better approximation to the Yukawa potential~\cite{Cassel:2009wt}. Furthermore, we have included an extra factor of $1/2$ to account for the fact that our annihilating particles are identical fermions, and thus only spin-singlet configurations contribute to this $s$-wave process (yielding a factor of $1/4$), but the overall cross section is increased by a factor of 2, as discussed in Ref.~\cite{Asadi:2016ybp}.

The experimental sensitivity to this cross section will depend on the final state to which the $h_D$ particles eventually decay, which in turn depends on $m_{h_D}$ and whether $h_D$ mixes with the SM-Higgs. However, in general hadronic decays will dominate the signal (due to the larger number of hadronic degrees of freedom), and the photon spectra from decays to different quark species are rather similar, as they arise largely from the decays of neutral pions produced in hadronic showers~\cite{Cirelli:2010xx}. 
Thus, we can estimate the sensitivity of indirect detection by examining the constraints set by assuming a $b\bar{b}$ final state.

In the left panel of Fig.~\ref{fig:indirect1} we show limits on the annihilation cross section to $b\bar{b}$ for Majorana DM from the \textit{Fermi}~\cite{Fermi-LAT:2016uux} and H.E.S.S.~\cite{Abdallah:2016ygi} gamma-ray telescopes, and sample results for the predicted annihilation rate from our two models. 
The H.E.S.S. limit, which dominates for DM masses above 1\,TeV, is based on a study of the region within 300\,pc of the Galactic Center, and assumes an Einasto density profile for the dark matter; if the Milky Way possesses a large core, these limits might be substantially weakened.
The \textit{Fermi} limits are based on a study of Milky Way dwarf spheroidal galaxies. The intermediate step of light mediator production will further broaden the photon spectrum, but Ref.~\cite{Elor2015a} demonstrated that the effect on the constraints is modest for hadronic final states where the spectrum is already quite broad. Thus to obtain an estimate of the constraints, we simply adopt the cross section limits for annihilation to $b\bar{b}$. We compare the maximum allowed cross section $\langle \sigma v_\text{rel} \rangle_\text{max}$ to the predicted cross section scaled by the fraction of DM in the $\chi_1$ state, $\langle \sigma v_\text{rel} \rangle (\Omega_{\chi_1} h^2/0.1198)^2$; examples for the pseudo-Dirac model with $y_D=2.5$ and the triple-Higgs model with $\alpha_D=3.0$ are shown in Fig.~\ref{fig:indirect1}, both for $m_V(m_8)=50\,$GeV.

In the left panel of Fig.~\ref{fig:indirect2}, we plot the regions in $m_\chi-m_{h_D}$ plane where bound states exist, the universe is not overclosed, and indirect limits are not violated. 
We see that there are almost no indirect constraints for DM masses below a few TeV and $m_{h_D}$ larger than the binding energy (when $m_{h_D}$ is below the binding energy, there remain allowed regions, but they must be chosen to avoid resonant Sommerfeld enhancement). We also plot the regions allowed by indirect detection bounds if a non-thermal history is assumed to ensure that $\chi_1$ constitutes 100\% of the DM, with $\Omega_\chi h^2 = 0.1198$. In this case, the indirect constraints are much more stringent, but the bulk of the region where $m_{h_D}$ exceeds the binding energy is still unconstrained.

\subsubsection{Triple-Higgs Model}

If the vector bosons other than $V^8$ are all at a heavy mass scale, then the dominant DM annihilation process (not involving bound states) both during freezeout and in the present day is tree-level annihilation to two $V^8$ bosons. This channel is available for $\bar{\chi}_i \chi_i$, where $i=1,2,3$; if $\sigma_i$ denotes the cross section for $\bar{\chi}_i \chi_i \rightarrow V^8 V^8$, then we have:
\begin{align} 
	\sigma_1 v_\text{rel} & = \sigma_2 v_\text{rel} = \frac{\pi (\alpha_D/12)^2}{m_\chi^2} = \frac{1}{16} \sigma_3 v_\text{rel}.
\end{align}
This channel furthermore experiences an attractive $s$-wave Sommerfeld enhancement, which for purposes of this estimate we approximate using Eq.~(\ref{eq:sommerfeld}).

Potential exchanges of $V^8$ bosons, which have large rates compared to processes involving the heavier gauge bosons, do not couple the $\bar{\chi}_i \chi_i$ and  $\bar{\chi}_j \chi_j$ states for $i \ne j$. Likewise, there is no (tree level) coannihilation to the $V^8 V^8$ final state. Thus, we can treat the $\chi_i$ species as evolving independently from each other, annihilating only with their own antiparticles, each experiencing its own long-range attractive Yukawa potential due to $V^8$ exchange. The effective couplings are $\alpha_D/12$ for $\chi_1$ and $\chi_2$ and $\alpha_D/3$ for $\chi_3$.

However, one important question is whether the different $\chi_i$ fields truly evolve independently, and in particular, whether decays and scatterings that interconvert between the $\chi_i$ states are rapid enough to keep the various state populations coupled during freezeout. An example process is $\bar{\chi}_1 \chi_1 \leftrightarrow \bar{\chi}_3 \chi_3$ scattering via $t$-channel $V^4$or $V^5$ exchange (see Appendix~\ref{app:triplehiggs}). As all such processes involve the heavier gauge bosons, they are slow compared to annihilation into a $V^8 V^8$ final state near the time of freezeout. For the $\overline{\chi}_1 \chi_1 \to \overline{\chi}_3 \chi_3$ process, the cross section for this scattering process is approximately $\sigma_{\chi_1 \overline{\chi}_1 \to \chi_3 \overline{\chi}_3} \sim \alpha_D^2 m_\chi^2/M^4$ where $M\sim m_{4,5}$. Compared with $\overline{\chi}_i \chi_i \to V^8 V^8\sim\alpha^2_D/m^2_\chi$, the rate of processes that scatter one type of fermion into another is suppressed by a factor of $\sim m_\chi^4/M^4$. Thus, the process $\chi_1\bar{\chi}_1 \to \chi_3\bar{\chi}_3$ freezes out before the $\chi_1\bar{\chi}_1 \to V^8V^8$ and is therefore not relevant to determining the relic abundance. 
This estimate assumes that all of the gluons other than $V^8$ are more massive than the DM; if this assumption breaks down, the three dark-matter-like populations will no longer evolve independently, and freezeout will be modified.

Under this assumption, we solve separate Boltzmann equations for each of the $\chi_i$ species (accompanied by their antiparticles), and require that the resulting mass density $2 (m_{\chi_1} n_{\chi_1} + m_{\chi_2}  n_{\chi_2} + m_{\chi_3} n_{\chi_3})$ matches the cosmological density of dark matter. The masses of the three states are assumed to be equal, with mass splittings small compared to the temperature at freezeout. The greater annihilation rate of $\bar{\chi}_3 \chi_3$ causes its abundance to be depleted faster than $\bar{\chi}_1 \chi_1$ and $\bar{\chi}_2 \chi_2$.

To estimate the late-time indirect detection limits, we proceed as for the pseudo-Dirac case above, and present our results in the right-hand panels of Figs.~\ref{fig:indirect1}-\ref{fig:indirect2}. The allowed cross section for DM annihilation is doubled as the DM $\chi_1$ is a Dirac fermion in this case. Since there is an unsuppressed $s$-wave annihilation channel, there are useful constraints from indirect detection even when radiative bound-state formation is kinematically forbidden. To estimate the contribution from bound-state formation, we numerically calculate the cross section for capture into the ground state by dipole photon emission, and also add the contribution from an $s$-wave initial state transitioning into the first excited state by emission of a dipole photon. The former process dominates when the mediator mass can be neglected \cite{Asadi:2016ybp}, but is suppressed in the very-low-velocity regime as it corresponds to a $p$-wave initial state \cite{An:2016gad}; thus we add the latter process to properly include the leading contribution at very small velocities. We follow the numerical method described in Ref.~\cite{Asadi:2016ybp}.

Note that in this case, the scenario where $\chi_1$ constitutes 100\% of the DM at late times is essentially completely excluded by indirect detection, for $\alpha_D=3.0$ and $m_\chi$ below 10 TeV; such a scenario requires a non-thermal origin for the DM, as the annihilation cross section is well above the thermal value and would deplete the DM density efficiently during freezeout. If non-thermal processes produce more DM at late times, then the large bare annihilation cross section and accompanying Sommerfeld enhancement (and possibly radiative bound-state formation) gives rise to very strong indirect detection signals, as shown in the right panel of Fig.~\ref{fig:indirect1}.

Both the overdepletion of the DM density and the large direct detection signals may be avoided if the Dirac-fermion DM possesses some tiny asymmetry, similar to the baryon asymmetry of the SM. The large annihilation cross sections found in these models can readily deplete the DM abundance to the point where the asymmetry sets the residual relic density, and then the indirect-detection signals are suppressed by the absence of the symmetric component. Note that if no such asymmetry is present, indirect detection limits may also pose challenges for sub-GeV DM and mediators as studied by Ref.~\cite{An:2015pva}; thermal relic DM can be quite generically excluded for sub-GeV mediators and sub-TeV DM \cite{Cirelli:2016rnw}.

This behavior does not occur for the pseudo-Dirac model because the principal annihilation channel is $p$-wave suppressed; this both makes it possible for TeV-scale $\chi_1$ particles to constitute 100\% of the DM with a thermal history, and ensures that large regions of parameter space remain that are not excluded by current indirect detection bounds (although bound state formation can provide indirect detection signals, as in Ref.~\cite{An:2016kie}).

\section{Discussion}
\label{sec:dis}

The resonant production of dark sector bound states at the LHC can be an important complementary search channel to the missing energy and mediator resonance searches. 
Unlike a mediator resonance search, a bound state resonance search directly probes the properties of the DM, and can be more effective when the mediator decays primarily to invisible DM particles. In addition, a bound state resonance search can be more sensitive than a missing energy search strategy at high DM masses.

We have studied the general features of models that can be probed by bound state resonance searches at the LHC while remaining consistent with other powerful experimental constraints. These models generally require a sufficiently strong coupling to a light mediator that can support the bound state, and a heavy mediator that couples the dark sector to the SM. We also carefully take into account the mixing between the heavy mediator and the dark matter bound state. Bound state decays into the light mediator should also be suppressed to allow for a significant partial width into SM particles. 

These requirements must be reconciled with constraints from both direct and indirect detection experiments. Spin-independent direct detection cross sections can be suppressed by having only loop-level interactions between the dark sector and nucleons, which can result from an off-diagonal coupling between the SM and two DM states with a mass splitting between them. Constraints from gamma-ray experiments and overclosure must be carefully considered, taking into account Sommerfeld enhancement due to the presence of a light mediator and radiative bound state formation both during freeze-out and in the present day. 

The SU(2)$_L$ minimal DM models possess many of the properties that we have discussed above, but pure wino or higgsino DM chargino bound states have a production cross section that lies well below the sensitivity of dilepton resonance searches, although DM particles in larger representations of SU(2)$_L$ with a large electric charge forming a deeply-bound electromagnetic bound state can potentially be discovered. 

We propose two dark sector models with kinetic mixing into the SM that contain bound states that can be probed effectively through bound state resonance searches at the LHC, while remaining consistent with direct and indirect detection constraints. The pseudo-Dirac model contains two Weyl fermions with a small mass splitting between them, capable of forming bound states through a light Higgs mediator, while the triple Higgs model is an SU(3) gauge theory with a single Dirac fermion in the fundamental representation, with all of the properties required for a viable model being generated by symmetry breaking of the gauge group. We study the LHC phenomenology of these models and find that bound states searches are complementary to both missing energy and vector mediator resonance searches, and are particularly powerful at high DM masses. A simple rescaling of our constraints indicates that future  27\,TeV or 100\,TeV $pp$ colliders could potentially probe DM bound states with masses in the $\mathcal{O}(10)$ TeV range.

We find that these models naturally avoid overclosure of the universe, and broad swaths of parameter space exist where they also evade limits from indirect detection searches under the assumption of a thermal history, despite the presence of the bound state implying model-independent large enhancements to the low-velocity annihilation rate. The indirect limits are most easily satisfied when radiative capture to the bound state in the local DM halo is kinematically forbidden, because the mass of the mediator supporting the bound state exceeds the binding energy. 

If the bound-state-forming species is required to constitute 100\% of the DM, through a non-thermal history, but is symmetric in the present day, then the indirect searches are sufficient to rule out almost all of the parameter space of interest in the LHC bound-state resonance search for the triple Higgs model; the pseudo-Dirac model evades this fate through a late-time velocity suppression of its annihilation rate. Where a DM species has a greater-than-thermal annihilation cross section but still constitutes 100\% of the observed DM density, a viable model that can be first detected by resonance searches at the LHC should possess some suppression to the annihilation cross section at late times, due e.g. to a dominant $p$-wave annihilation channel or a small primordial asymmetry.

To summarize, dark sectors with bound states can be probed at the LHC through resonance decays to SM particles. Models with multiple force carriers and DM-like states, where the DM scatters inelastically off SM quarks at tree-level, can naturally give rise to a sufficiently large production cross section while evading direct detection constraints. The presence of a light mediator, needed to support the bound state, modifies freezeout and leads to stringent indirect detection limits; however, these constraints leave a wide region of parameter space open, while suggesting a preferred mass spectrum where the mediator mass exceeds the binding energy. DM models with bound states possess a rich phenomenology, allowing complementary probes from many different experimental directions.

\section*{Acknowledgements}

The authors would like to thank Pouya Asadi, Spencer Chang, Tim Cohen, Yevgeny Kats, Graham Kribs, Marat Freytsis, Iain Stewart, Matthew Strassler and Wei Xue for helpful discussions, and Tim Cohen for comments on the draft. We particularly thank Yevgeny Kats for pointing out the importance of mixing between bound states and mediators.
Feynman diagrams in this paper were generated with TikZ-Feynman~\cite{Ellis:2016jkw}, and Feynman amplitude calculations performed with FeynCalc~\cite{Mertig:1990an,Shtabovenko:2016sxi}.
This work was supported in part by the Office of High Energy Physics of the U.S. Department of Energy under grant Contract Number DE-SC00012567. GE is supported by a National Science Foundation LHC Theory Initiative Postdoctoral Fellowship, under the grant number PHY-0969510, and by the U.S. Department of Energy, under grant numbers DE-SC001819 and DE-SC0011637. HL is supported by the MIT Research Support Committee. TRS and YS were also supported by the U.S. Department of Energy under grant Contract Numbers DE-SC0013999 and DE-SC-00015476 respectively. 

\appendix

\section{Bound State Matrix Elements}
\label{sec:BoundStateTheory}

The Feynman diagram for vector-mediated bound state formation from $q \overline{q}$ interactions is shown in Fig.~\ref{fig:feynmanMonoXAndResonance}. To calculate the matrix element associated with bound state formation, we adopt the Non-Relativistic QCD~(NRQCD) formalism used in \cite{Kats:2009bv}, and see also  \cite{Petrelli:1997ge,An:2015pva}.
Let the amplitude associated with the production of a pair of free dark matter particles from $q \overline{q}$, with the external spinors for the DM fermions amputated, be denoted by $\mathcal{A}_0$. The corresponding amplitude for the formation or decay of a non-relativistic, spin-triplet, $L=0$ bound state $\mathcal{B}$ is then given by   
\begin{alignat}{1}
	\mathcal{A}_{\mathcal{B}} = \epsilon_\alpha \text{Tr} [\Pi^\alpha \mathcal{A}_0],
\end{alignat}
where $\epsilon_\alpha$ is the polarization vector associated with a massive spin-1 particle. The mass of $\mathcal{B}$ is taken to be $m_{\mathcal{B}} = 2 m_\chi$. $\Pi^\alpha$ is a projection operator that gives the correct final spin-state of the DM particles and accounts for the wavefunction overlap associated with the bound state, given by
\begin{alignat}{1}
	\Pi^\alpha = \frac{\psi^*(0)}{\sqrt{8 m_\chi^3}} (\slashed{p} - m_\chi) \gamma^\alpha (\slashed{p} + m_\chi),
\end{alignat}
where $p$ is the final 4-momentum of each of the two DM particles, and $\psi(0)$ is taken to be the hydrogen-like ground state wavefunction, 
\begin{alignat}{1}
	|\psi(0)|^2 = \frac{\alpha_\mathcal{B}^3 m_\chi^3}{8\pi},
	\label{eqn:psi2}
\end{alignat}
where $\alpha_\mathcal{B}$ is the coupling constant between $\chi$ and the bound state mediator. We stress here that in a generic model with multiple dark sector mediators, the mediator that produces the bound state need not be the same as the mediator that binds the dark matter (see Fig.~\ref{fig:feynmanMonoXAndResonance}). 

\section{Details of Kinetic Mixing}
\label{app:kineticMixing}

The kinetic mixing term in both the pseudo-Dirac U(1) model and the triple Higgs SU(3)$_D$ model is of the form
\begin{alignat}{1}
	\mathcal{L} \supset -\frac{1}{4} V_{\mu\nu}V^{\mu\nu} -\frac{\epsilon}{2} V_{\mu\nu} B^{\mu\nu} - \frac{1}{4} B_{\mu\nu} B^{\mu\nu} \, ,
\end{alignat}
where $V_{\mu\nu}$ is the dark sector gauge field strength. To diagonalize this, we define the new field $B'_\mu = B_\mu + \epsilon V_\mu$, and thus $B'_{\mu\nu} = B_{\mu\nu} + \epsilon V_{\mu\nu}^{\text{ab.}}$, where ab. indicates the abelian part of the field strength tensor. All terms of $\mathcal{O}(\epsilon^2)$ are neglected here. Then
\begin{alignat}{1}
	\mathcal{L} \supset -\frac{1}{4} V_{\mu\nu} V^{\mu\nu} - \frac{1}{4} B'_{\mu\nu} B'^{\mu\nu} - \frac{\epsilon g_D}{2} f^{abc} V_\mu^b V_\nu^c B'^{\mu\nu}  \, ,
\end{alignat}
The last term is an additional interaction that is unimportant in the diagonalization. In terms of the new field, we have the SM fields
\begin{alignat}{1}
	Z_\mu 
	&= c_W W^3_\mu - s_W B_\mu 
	= c_W W^3_\mu - s_W(B'_\mu - \epsilon V_\mu) \nonumber \\
	&= Z'_\mu + \epsilon s_W V_\mu, \nonumber \\
	A_\mu 
	&= s_W W^3_\mu + c_W B_\mu 
	= s_W W^3_\mu + c_W(B'_\mu - \epsilon V_\mu) \nonumber \\
	&= A'_\mu - \epsilon c_W V_\mu,
\end{alignat}
where we have defined $Z_\mu' = c_W W_\mu^3 - s_W B'_\mu$ and $A_\mu' = s_W W_\mu^3 + c_W B'_\mu$. 
In terms of these new fields, the mass terms for $Z$ and $V$ are:
\begin{alignat}{1}
	\mathcal{L} 
	&\supset 
	\frac{1}{2} m_Z^2 Z_\mu Z^\mu + \frac{1}{2} g^2 v^2 V_\mu V^\mu \nonumber \\
	&= \frac{1}{2} m_Z^2 Z'_\mu Z'^\mu + \epsilon s_W m_Z^2 Z_\mu' V^\mu  +  \frac{1}{2} g^2 v^2 V_\mu V^\mu.
\end{alignat}
Diagonalizing this and defining $r \equiv m_Z/m_V$ (we neglect the $\mathcal{O}(\epsilon^2)$ shift in the masses), we get the following mass eigenstates to first order in $\epsilon$ (marked by tildes):
\begin{alignat}{1}
	\tilde{Z}_\mu &= Z'_\mu - r^2 \frac{\epsilon s_W}{1 - r^2} V_\mu \, , \nonumber \\
	\tilde{V}_\mu &= V_\mu + r^2 \frac{\epsilon s_W}{1 - r^2} Z'_\mu \, ,
\end{alignat}
with $A'$ being the massless mode. 
With this, we see that the original SM fields become
\begin{alignat}{1}
	Z_\mu &= \tilde{Z}_\mu + \frac{\epsilon s_W}{1 - r^2} \tilde{V}_\mu, \\
	A_\mu &= A'_\mu - \epsilon c_W \tilde{V}_\mu,
\end{alignat}
and
\begin{alignat}{1}
    V_\mu &= \tilde{V}_\mu - r^2 \frac{\epsilon s_W}{1 - r^2} \tilde{Z}_\mu.
\end{alignat}
This is the result shown in Eq.~(\ref{eqn:currents}).

\section{Details of the Triple Higgs Model}
\label{app:triplehiggs}

\begin{table}
    \begin{center}
    \begin{tabular}{cc}
        \begin{tikzpicture}
            \begin{feynman}
                \vertex (chi2) {\(\chi_2\)};
                \vertex[right=1.4cm of chi2] (int);
                \vertex[above right=1cm and 1cm of int] (gluon) {\(V^{1,2}\)};
                \vertex[below right=1cm and 1cm of int] (chi1) {\(\chi_1\)};
                \diagram*{
                    (chi2) -- [fermion] (int);
                    (int) -- [boson] (gluon);
                    (int) -- [fermion] (chi1);
                };
            \end{feynman}
        \end{tikzpicture} &
        \begin{tikzpicture}
            \begin{feynman}
                \vertex (chi3) {\(\chi_3\)};
                \vertex[right=1.4cm of chi3] (int);
                \vertex[above right=1cm and 1cm of int] (gluon) {\(V^{4,5}\)};
                \vertex[below right=1cm and 1cm of int] (chi1) {\(\chi_1\)};
                \diagram*{
                    (chi3) -- [fermion] (int);
                    (int) -- [boson] (gluon);
                    (int) -- [fermion] (chi1);
                };
            \end{feynman}
        \end{tikzpicture} \\
        \begin{tikzpicture}
          \begin{feynman}
            \vertex (chi3) {\(\chi_3\)};
                \vertex[right=1.4cm of chi3] (int);
                \vertex[above right=1cm and 1cm of int] (gluon) {\(V^{6,7}\)};
                \vertex[below right=1cm and 1cm of int] (chi2) {\(\chi_2\)};
                \diagram*{
                    (chi3) -- [fermion] (int);
                    (int) -- [boson] (gluon);
                    (int) -- [fermion] (chi2);
                };
          \end{feynman} 
        \end{tikzpicture} & 
        \begin{tikzpicture}
          \begin{feynman}
            \vertex (chi123) {\(\chi_{1,2}\)};
                \vertex[right=1.8cm of chi123] (int);
                \vertex[above right=1cm and 1cm of int] (gluon) {\(V^{3}\)};
                \vertex[below right=1.cm and 0.8cm of int] (chiout) {\(\chi_{1,2}\)};
                \diagram*{
                    (chi123) -- [fermion] (int);
                    (int) -- [boson] (gluon);
                    (int) -- [fermion] (chiout);
                };
          \end{feynman} 
        \end{tikzpicture} \\
        \begin{tikzpicture}
          \begin{feynman}
            \vertex (chi123) {\(\chi_{1,2,3}\)};
                \vertex[right=1.8cm of chi123] (int);
                \vertex[above right=1cm and 1cm of int] (gluon) {\(V^{8}\)};
                \vertex[below right=1.cm and 0.8cm of int] (chiout) {\(\chi_{1,2,3}\)};
                \diagram*{
                    (chi123) -- [fermion] (int);
                    (int) -- [boson] (gluon);
                    (int) -- [fermion] (chiout);
                };
          \end{feynman} 
        \end{tikzpicture}
    \end{tabular}
    \end{center}
    
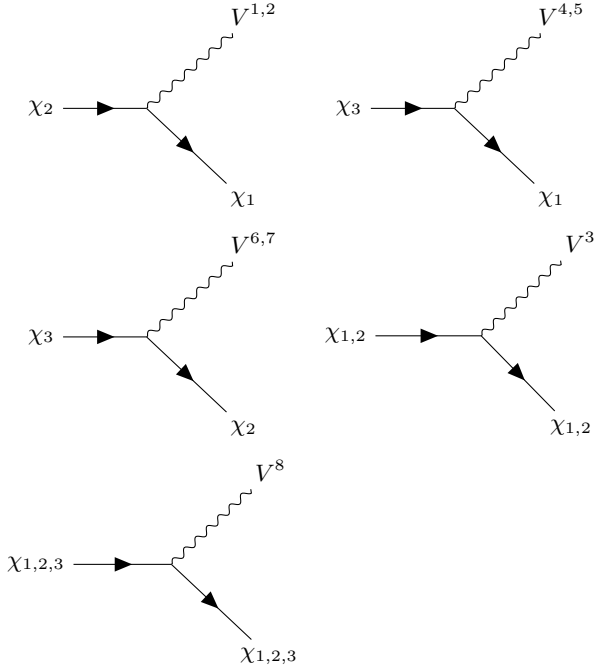
\captionof{figure}{Gauge-fermion-fermion vertices for the triple Higgs model.}
    \label{fig:feynmanTripleHiggs}
\end{table}

Figure \ref{fig:feynmanTripleHiggs} shows all the gauge-fermion-fermion interaction vertices in the triple Higgs model. These vertices are important in the discussion of the stability of the non-DM fermions and thermal freeze-out.

In the triple Higgs model of Sec.~\ref{sec:TripleHiggs}, $H_1$ and $H_2$ acquire a VEV $v_1$ and $v_2$ in the 1- and 2-directions respectively, at a scale above the bare Dirac mass $m_\chi$. This breaks the gauge symmetry down to a residual U(1). Subsequently, this remaining symmetry group is broken at a lower scale than $m_\chi$ by $H_8$, which acquires a VEV given by
\begin{alignat}{1}
	\langle H_8 \rangle = \begin{pmatrix}
	v_8 \cos \theta \\ 0 \\ v_8 \sin \theta
\end{pmatrix},
\end{alignat}
with $v_8 \ll m_\chi$. This symmetry breaking pattern can be achieved with the following Higgs potential, which obeys the $\mathbb{Z}_2$ symmetry mentioned above: 
\begin{multline}
	V(H_1, H_2, H_8) = \sum_{i=1,2} \lambda_i (H_i^a H_i^a - v_i^2)^2 + \lambda_8 (H_8^\dagger H_8 - v_8^2)^2 \\
	+ \lambda_{12} (H_1^a H_2^a)^2 + \frac{1}{\Lambda_H^2} \sum_{i=1,2} (H_8^\dagger \tau^a H_8 H_i^a)^2.
\end{multline}
The last two terms forbid a VEV in the second component of $H_8$, which would make $H_8^\dagger \tau^a H_8$ non-zero for $a = 1$. This symmetry breaking pattern produces the mass hierarchy for the dark gluons and the mass splitting between the Dirac fields in $\chi$, which were discussed in the main text.  

After symmetry breaking, the kinetic mixing terms with the SM become
\begin{multline}
	\mathcal{L}_{\text{mix}} = -\frac{\epsilon}{2} V_{\mu\nu}^1 B^{\mu\nu} - \frac{\epsilon_8}{2} \left[ \cos^2 \theta \, V_{\mu\nu}^3 + \sin 2 \theta \, V_{\mu\nu}^4  \right] B^{\mu\nu} \\
	- \frac{\epsilon_8}{2 \sqrt{3}} \left( \cos^2 \theta - 2 \sin^2 \theta \right) V_{\mu\nu}^8 B^{\mu\nu},
\end{multline}
where we have defined $\epsilon_8/2 \equiv v_8^2/\Lambda_8^2$, taking $\epsilon_8 \ll \epsilon \ll 1$. The first term represents the kinetic mixing between dark sector and the SM discussed in the main text, while the remaining mixing terms are highly suppressed but non-zero for generic values of $\theta$: their existence guarantees that the gluons $V^3$, $V^4$ and $V^8$ decay to SM particles over cosmological timescales. 

With the $m_8 < 2m_\chi < m_{1,\cdots,7}$ mass hierarchy, the only stable dark sector particle is $\chi_1$, since we can assign a conserved dark baryon number to the fermions and $\chi_1$ is the lightest dark fermion. The heavy gluons $V^i$ for $i = 1, \cdots, 7$ can decay into a pair of dark fermions since their masses exceed $2m_\chi$. Decays into a pair of dark fermions for $V^i$ with $i = 2, 5, 6$ and 7 occur promptly for an $\mathcal{O}(1)$ coupling $g_D$. For $V^8$, which mixes directly into the SM and can decay into a pair of SM fermions, its lifetime is approximately
\begin{alignat}{1}
	\tau_{8} \sim \left(\frac{10^{-11}}{\epsilon_8}\right)^2 \left(\frac{100 \text{ MeV}}{m_8}\right) \times 1 \text{ s}.
\end{alignat}
For the fermions, both the $\chi_2$ and $\chi_3$ fermion can decay to $ \chi_1 f \overline{f} $ where $f$ is an SM fermion through an off-shell $V^1$ and $V^4$ respectively. The decay lifetime for the heavier dark fermions are approximately
\begin{alignat}{1}
	\tau_{\chi_2} &\sim \left(\frac{0.1}{\epsilon}\right)^2 \left(\frac{m_1}{\text{TeV}}\right) \left(\frac{\text{TeV}}{m_\chi}\right)^5 \left(\frac{10^{-6}}{\Delta m_\chi/m_\chi}\right)^5 \times 1 \text{ s}, \nonumber \\
	\tau_{\chi_3} &\sim \left(\frac{0.1}{\epsilon_8}\right)^2 \left(\frac{m_4}{\text{TeV}}\right) \left(\frac{\text{TeV}}{m_\chi}\right)^5 \left(\frac{10^{-6}}{\Delta m_\chi/m_\chi}\right)^5 \times 1 \text{ s}.
\end{alignat}
where $\Delta m_\chi \equiv v_8^2/2\Lambda_m$ is the mass splitting between $\chi_2$, $\chi_3$ and $\chi_1$. For any reasonable choice of parameters considered in this paper, it is clear that $V^8$, $\chi_2$ and $\chi_3$ are all unstable on cosmological timescales, and that the only dark matter component in the present day is $\chi_1$. 

\section{Decay widths}
\label{app:widths}
Table~\ref{tab:BDecayWidths} shows the perturbative partial decay widths of the dark sector bound state $\cB$ in both models into SM final states, through mixing with the SM mediator $V$. 
The bound state wavefunction is given in Eq. (\ref{eqn:psi02}). 
In addition, Table~\ref{tab:VDecayWidths} shows the perturbative partial decay widths of the SM mediator $V$ into all possible final states.

\renewcommand{\arraystretch}{3}

\setlength{\tabcolsep}{15pt}

\begin{table*}
\begin{tabular}{c c}
\toprule

Decay Process & $\frac{\Gamma(\mathcal{B} \to X)}{\alpha_D |\psi(0)|^2}$ \\
\hline
$\mathcal{B} \to f \overline{f}$ & $\frac{16 \pi N_m N_c r^4 \alpha \epsilon^2 m_\chi^2 \left[ \left(c_W^2 Q \left(m_Z^2 - 4m_\chi^2 \right) + 4 g_V m_\chi^2 \right)^2 + 16 g_A^2 m_\chi^4 + r^2 \Gamma_V^2 m_Z^2(g_V^2 + g_A^2)/(1-r^2)^2 \right]}{3 c_W^2 \left(m_Z^2 - 4m_\chi^2\right)^2 \left[ (m_Z^2 - 4 r^2 m_\chi^2)^2 + r^2 \Gamma_V^2 m_Z^2 \right]}$ \\
$\mathcal{B} \to W^+W^-$ & $\frac{4 \pi N_m r^4 c_W^2 \alpha \epsilon^2 \left(m_\chi^2 - m_W^2 \right)^{3/2} \left(4m_\chi^4 + 20 m_\chi^2 m_W^2 + 3m_W^4 \right) \left[ m_Z^4 + r^2 \Gamma_V^2 m_Z^2/(1 - r^2)^2\right] }{3 m_\chi m_W^4 \left(m_Z^2 - 4m_\chi^2 \right)^2 \left[(m_Z^2 - 4 r^2 m_\chi^2)^2 + r^2 \Gamma_V^2 m_Z^2 \right]}$ \\
$\mathcal{B} \to Z h^0$ & $\frac{\pi N_m r^4 \alpha \epsilon^2 \left[m_Z^4 + 2 m_Z^2 \left(20 m_\chi^2 - m_H^2 \right) + \left(m_H^2 - 4m_\chi^2\right)^2 \right] \sqrt{m_Z^4 - 2m_Z^2 \left(m_H^2 + 4m_\chi^2 \right) + \left(m_H^2 - 4m_\chi^2\right)^2} \left[16 m_\chi^4 + r^2 \Gamma_V^2 m_Z^2/(1 - r^2)^2 \right] }{192 c_W^2 m_\chi^4 \left(m_Z^2 - 4m_\chi^2 \right)^2 \left[ (m_Z^2 - 4 r^2 m_\chi^2)^2 + r^2 \Gamma_V^2 m_Z^2 \right]}$ \\
$\mathcal{B}_{\text{pD}} \to V h_D$ & $\frac{\pi \alpha_D \left[m_V^4 + 2 m_V^2 \left(20 m_\chi^2 - m_{h_D}^2 \right) + \left(m_{h_D}^2 - 4 m_\chi^2 \right)^2 \right] \sqrt{m_V^4 - 2 m_V^2 \left(m_{h_D}^2 + 4m_\chi^2 \right) + \left(m_{h_D}^2 - 4 m_\chi^2 \right)^2 }}{3 m_\chi^4 \left(m_V^2 - 4 m_\chi^2 \right)^2}$ \\
\botrule
\end{tabular}
\caption{Table of perturbative partial widths for the bound state $\mathcal{B}$ in both the dark sector models. $N_m = 4$ for the pseudo-Dirac model, and $N_m = 1$ for the triple Higgs model: this factor accounts for differences in the type of fermion in each theory, as well as the value of the coupling between the DM and the SM mediator. $N_c = 3$ for quarks and 1 otherwise, $g_V = g_{V,Z} \equiv \{0.25, -0.0189, 0.0959, -0.1730 \}$ and $g_A = g_{A,Z} \equiv \{0.25, -0.25, 0.25, -0.25\}$ are the vector and axial couplings to the $Z$-boson for $\{\nu_e, e, u, d\}$ and for the other 2 generations respectively. $Q$ is the electric charge of each species, and $\alpha$ electromagnetic fine structure constant; $\Gamma_V$ is the width of the SM mediator in each model ($V$ in the pseudo-Dirac model, $V^1$ in the triple Higgs), $m_H$ is the mass of the SM Higgs, and $m_{h_D}$ is the mass of the dark sector Higgs in the pseudo-Dirac model. The last expression is only applicable to the pseudo-Dirac model.}
\label{tab:BDecayWidths}
\end{table*}

\renewcommand{\arraystretch}{3}

\setlength{\tabcolsep}{15pt}

\begin{table*}
\begin{tabular}{c c}
\toprule

Decay Process & Partial Width \\
\hline
$V \to \chi \overline{\chi}$ & $\frac{g_D^2}{12 \pi} \sqrt{m_V^2 - 4 m_\chi^2}$ \\
$V \to f \overline{f}$ & $\frac{N_c e^2 \epsilon^2}{12 \pi c_W^2(1 - r^2)^2}  \sqrt{m_V^2 - 4 m_f^2} \left[(g_V^2 + g_A^2) + \frac{2m_f^2}{m_V^2} (g_V^2 - 2g_A^2)\right]$ \\
$V \to W^+ W^-$ & $\frac{\epsilon^2 e^2 c_W^2 r^4 m_V}{192 \pi (1 - r^2)^2} \frac{(1 - 4x^2)^{3/2}}{x^4} (1 + 20x^2 + 12 x^4)$ \\
$V \to Zh^0$ & $\frac{\epsilon^2 e^2 m_V}{192 \pi c_W^2 r^2 (r^2 - 1)^2} \sqrt{(y^2 - 1)^2 r^4 - 2r^2(y^2 + 1) + 1} \left[r^4(y^2 - 1)^2 - 2r^2(y^2 - 5) + 1 \right]$ \\
\botrule
\end{tabular}
\caption{Table of perturbative partial widths for the SM mediator $V$ for both the dark sector models. $N_c = 3$ for quarks and 1 otherwise. $g_V = g_{V,Z} - c_W^2(1 - r^2)Q$ and $g_A = g_{A,Z}$, where $ g_{V,Z} \equiv \{0.25, -0.0189, 0.0959, -0.1730 \}$ and $g_{A,Z} \equiv \{0.25, -0.25, 0.25, -0.25\}$ are the vector and axial couplings to the $Z$-boson for $\{\nu_e, e, u, d\}$ and for the other 2 generations respectively, and $Q$ is the electric charge of the fermion. $r \equiv m_Z/m_V$, $x \equiv m_W/m_V$ and $y \equiv m_H/m_V$.}
\label{tab:VDecayWidths}
\end{table*}

\section{Bound State Formation via Initial/Final State Radiation}

In addition to the resonant formation process that has been our main focus in the body of this work, bound states can also form in conjunction with radiation of other particles in the initial or final state. This process is very important in the context of electron accelerators where the center-of-mass energy of the colliding particles is fixed and does not overlap the bound state resonance (as discussed \textit{e.g.} in Refs.~\cite{Brodsky:2009gx,An:2015pva}), and so the resonant signal is absent. 

This process could also be critical if the decays of spin-0 bound states were much more observable than those of spin-1 bound states, and the mediator with the SM were a vector (or if the spin-1 states were more observable and the mediator were a scalar); emission of additional particles would then allow the production of the rarer but more observable bound state. However, in the examples we have studied, the latter situation does not hold; indeed, the spin-0 $s$-wave bound states can generally decay into light mediators and are thus likely to be more difficult to detect than their spin-1 counterparts.

Since initial and final state radiation inevitably involves extra powers of the coupling relative to the resonant case, we expect processes of this type to be suppressed relative to the resonant production. However, one might wonder whether threshold enhancements to the production and interaction cross section for unbound but slow-moving DM particles, in the presence of a light mediator, could modify this conclusion and lead to a large contribution from the threshold region. 

Note that this is a very different parameter regime to that considered for muonium production in Ref.~\cite{Brodsky:2009gx} and for light darkonium production in Ref.~\cite{An:2015pva}, where the beam energy is presumed to be large relative to the resonance energy, and the extra particle(s) emitted as initial/final state radiation carry away much of the beam energy; it is more similar to the situation in indirect detection, where slow-moving DM particles may emit a light particle and radiatively capture into a bound state (see \textit{e.g.}~\cite{Pospelov:2008jd, MarchRussell:2008tu}). The rate for such radiative bound-state formation scales as $1/v$ close to threshold, for a massless mediator. However, we will show that in the case where the particles are produced near threshold and then form a bound state, this $1/v$ scaling is canceled out by the small phase space for the particle production near threshold.

Similar contributions to bound-state formation from soft gluon emission have been studied in the context of quarkonium formation using non-relativistic effective field theory techniques~\cite{Bodwin:1992qr, Bodwin:1994jh}. In that case, $p$-wave color-singlet quarkonia can be formed either directly or through an intermediate $s$-wave color-octet pair of heavy quarks; relatedly, the $s$-wave quarkonium state $|Q\bar{Q}\rangle$ can be described as having a small $\mathcal{O}(v^2)$ admixture of a Fock state $|Q\bar{Q} g\rangle$ containing an additional soft gluon. This approach suggests that the admixture term can be neglected to leading order when dealing with $s$-wave bound states, and should not experience large enhancements near threshold.

To see explicitly how this works in our case, note that we can write the matrix element for production of the bound state (plus a light mediator with momentum $\vec{l}$), via an intermediate state of two near-threshold (i.e. highly non-relativistic) but unbound DM particles, as:
\begin{align} 
	& i\mathcal{M}(i\rightarrow f) \nonumber \\
	& = \sqrt{\frac{2 M}{(2 m_\chi)(2 m_\chi)} } \int \frac{d^3 p_1}{(2\pi)^3}   \frac{2\pi \delta(E_\text{in} - E_{p_1} - E_{p_2})}{(2 E_{p_1})(2 E_{p_2})}  \nonumber \\ 
 	& \times \int \frac{d^3 a}{(2\pi)^3} \tilde{\psi}_{p_1,p_2}^*(\vec{a})i \bar{\mathcal{M}}(i \rightarrow \vec{a}_1, \vec{a}_2) \nonumber \\ 
	& \times   \int \frac{d^3 b}{(2\pi)^3} \tilde{\psi}_{p_1,p_2}(\vec{b})   \times \int \frac{d^3 q}{(2\pi)^3} \tilde{\psi}^*_B(\vec{q})  \nonumber \\
 	& \times  i\bar{\mathcal{M}}(\vec{b}_1, \vec{b}_2 \rightarrow \vec{q}_1 \vec{q}_2 \vec{l}).\label{eq:fsrmaster} 
\end{align}
 Here $\bar{\mathcal{M}}(i \rightarrow \vec{a}_1, \vec{a}_2)$ is the hard matrix element describing production of two free DM particles with momenta $\vec{a}_1, \vec{a}_2$ from the initial state $i$, and likewise $\bar{\mathcal{M}}(\vec{b}_1, \vec{b}_2 \rightarrow \vec{q}_1 \vec{q}_2 \vec{l})$ is the hard matrix element describing the radiation of a light mediator with momentum $\vec{l}$ from the DM-DM state with particle momenta $\vec{b}_1, \vec{b}_2$, to produce final-state DM particles with momenta $\vec{q}_1, \vec{q}_2$. The wavefunctions convert the plane-wave states to the full intermediate and final states accounting for potential effects. $\vec{p}_1$ and $\vec{p}_2$ act as labels on the intermediate state with momentum-space wavefunction $\tilde{\psi}_{p_1, p_2}$, describing the momenta of the constituent particles at large separation. $\tilde{\psi}_B$ denotes the momentum-space wavefunction of the bound state (which in principle is labeled by the quantum numbers $n, l, m$; we suppress these indices). $m_\chi$ is the DM mass and $M \approx 2 m_\chi$ is the bound-state mass.

In the non-relativistic limit where the potential is neglected, the leading-order matrix element for light vector boson radiation from one of a pair of heavy fermions (with gauge coupling $g_\mathcal{B}$ and fermion masses $m_1$, $m_2$) is given by:
\begin{align} 
 	& i\mathcal{M}(\vec{b}_1, \vec{b}_2 \rightarrow \vec{q}_1 \vec{q}_2 \vec{l}) \nonumber \\
	& = i g_\mathcal{B} \vec{\epsilon^*}(l) \cdot \left[ (\vec{b}_1 + \vec{q}_1) 2 m_2 (2\pi)^3 \delta^{(3)}(\vec{b}_2 - \vec{q}_2) \right. \nonumber \\
	& \left. - (\vec{b}_2 + \vec{q}_2) 2 m_1 (2\pi)^3 \delta^{(3)}(\vec{b}_1 - \vec{q}_1)  \right]. 
\end{align}

Inserting this expression into Eq.~\eqref{eq:fsrmaster}, setting the masses of the two heavy fermions equal, $m_1 = m_2 = m_\chi$, working in relative momentum coordinates, and choosing the center-of-mass frame, we obtain:
\begin{align} 
	& i\mathcal{M}(i\rightarrow f) \nonumber \\
 	& =2 i g_\mathcal{B} \sqrt{2 M} \vec{\epsilon^*}(l)  \cdot   \int \frac{d^3 p_1}{(2\pi)^3} \frac{2\pi \delta(E_\text{in} - E_{p_1} - E_{p_2})}{(2 E_{p_1})(2 E_{p_2})} \nonumber \\
 	& \times \left[\int \frac{d^3 a}{(2\pi)^3} i \bar{\mathcal{M}}(i \rightarrow \vec{a}_1, \vec{a}_2) \tilde{\psi}_{p_1,p_2}^*(\vec{a})  \right]  \nonumber \\
 	& \times \int \frac{d^3 q}{(2\pi)^3}        \tilde{\psi}^*_B(\vec{q}) \vec{q} \left(  \tilde{\psi}_{p_1,p_2}(\vec{q} + \vec{l}/2) +  \tilde{\psi}_{p_1,p_2}(\vec{q} - \vec{l}/2)  \right) .
\end{align}
 The integral over $d^3 q$ on the last line also appears in the matrix element for radiative bound state formation, and has been previously computed in the non-relativistic limit for massless vector mediators \cite{0953-4075-29-10-021,Asadi:2016ybp}. In the near-threshold regime, $l \lesssim \alpha^2 m_\chi$ (as the binding energy must provide the necessary energy to radiate the mediator), and the $l$-dependence of the integral can be neglected; in this case, the integral simply scales as $1/\sqrt{p}$, where $\vec{p} = (\vec{p}_1 - \vec{p}_2)/2$. (This factor, when squared, is responsible for the $1/v$ scaling of the radiative bound state formation cross section.)
 
 If we further suppose that the hard matrix element for production of the intermediate state from the initial state is independent of the final-state relative momentum $\vec{a}$, i.e. we can write $i \bar{\mathcal{M}}(i \rightarrow \vec{a}_1, \vec{a}_2) = i \bar{\mathcal{M}}(i \rightarrow \text{DM}, \text{DM})$  then the integral over $d^3 a$ simplifies to give $i \bar{\mathcal{M}}(i \rightarrow \text{DM}, \text{DM}) \psi^*_{p_1,p_2}(0)$, where $\psi$ denotes the position-space wavefunction. The wavefunction at the origin in a Coulomb-like potential scales as $\sqrt{\alpha_\mathcal{B} m_\chi/p}$ (\textit{e.g.}~\cite{0953-4075-29-10-021}), which yields the usual Sommerfeld enhancement when squared.
 
 Putting these pieces together and performing the phase-space integral over $\int d^3 p_1$, writing $E_{p_1} = E_{p_2} = \sqrt{m_\chi^2 + |\vec{p}|^2}$ since we are working in the COM frame, we find that (keeping only scaling relationships, dropping order-1 factors):
 \begin{align}  
 	& i\mathcal{M}(i\rightarrow f) \sim g_\mathcal{B} \sqrt{m_\chi} i \bar{\mathcal{M}}(i \rightarrow \text{DM}, \text{DM}) \nonumber \\
	& \times \frac{1}{E_\text{in}^2} \int \frac{d^3 p}{(2\pi)^3} 2\pi \delta\left(E_\text{in} - 2 \sqrt{m_\chi^2 + |\vec{p}|^2}\right) \sqrt{\frac{\alpha_\mathcal{B} m_\chi}{p}} \sqrt{\frac{1}{p}} \nonumber \\
 	& \sim \bar{\mathcal{M}}(i \rightarrow \text{DM}, \text{DM}) g_\mathcal{B} m_\chi \frac{\sqrt{\alpha_\mathcal{B}}}{E_\text{in}} \nonumber \\
 	& \sim  \bar{\mathcal{M}}(i \rightarrow \text{DM}, \text{DM}) \alpha_\mathcal{B} . 
\end{align}
 Note that as mentioned previously, the phase-space integral over the intermediate-state momentum $d^3 p$ has canceled out the $1/p$ scaling from the wavefunctions.
 
 Thus we see that the bound-state production cross section through this channel should scale as $|\bar{\mathcal{M}}(i \rightarrow \text{DM}, \text{DM})|^2 \alpha_\mathcal{B}^2$, multiplied by a 2-body phase space factor. Since the momenta in the final state are small, of order $l \sim \alpha_\mathcal{B}^2 m_\chi$, the overall scaling of the cross section with the couplings is $\alpha_\mathcal{B}^4 \times |\bar{\mathcal{M}}(i \rightarrow \text{DM}, \text{DM})|^2$.
 
 By comparison, the resonant production cross section scales as $|\bar{\mathcal{M}}(i \rightarrow \text{DM}, \text{DM})|^2 \alpha_\mathcal{B}^3$, where the $\alpha_\mathcal{B}$ dependence arises from the $\mathcal{B}$ wavefunction. Thus the rate to produce an extra light mediator by emission from a near-threshold intermediate state, in conjunction with the bound state formation, is suppressed by one power of $\alpha_\mathcal{B}$ overall. This is the same suppression one would naively expect for emission of a hard photon from the initial or final state, with no small phase-space factors or threshold enhancements. We self-consistently neglect all such diagrams in the body of this work.
 
 Here we have neglected the mediator mass $m_Y$ in estimating the scalings; in particular, the intermediate-state position-space wavefunction may be steeply peaked near the origin for special values of $m_Y$, corresponding to the presence of near-zero-energy bound states (\textit{e.g.}~\cite{Hisano:2004ds}). However, it seems likely that any apparent enhancement from this behavior can be reinterpreted as resonant capture into a near-zero-energy bound state, which is already accounted for in our formalism. We leave a more detailed study of the resonant regime to future work.

\bibliography{darkonium}

\end{document}